\documentclass[sigconf, nonacm]{acmart}
\usepackage{pvldb}

\AtBeginDocument{%
  }

\usepackage[most]{tcolorbox}

\usepackage{array}
\usepackage{svg}
\usepackage{subcaption}
\usepackage{tikz}
\usepackage{pgfplots}
\pgfplotsset{compat=1.18}
\usepackage{verbatim}
\usepackage{amsmath,bm}
\usepackage{enumitem}

\usepackage{multirow}

\usepackage{makecell}

\usepackage{tkz-euclide}
\usepackage{marginnote}

\usepackage{diagbox}
\usepackage{xcolor}
\usepackage{tabularx}
\usepackage{booktabs}
\usepackage[linesnumbered,ruled,vlined]{algorithm2e}

\lstdefinestyle{umma}{
  basicstyle=\ttfamily\scriptsize,   %
  keywordstyle=\bfseries,              %
  commentstyle=\color{gray},           %
  stringstyle=\color{teal},            %
  showstringspaces=false,
  columns=fullflexible,                %
  keepspaces=true,                     %
  frame=single,                        %
  rulecolor=\color{black!30},          %
  frameround=tttt,
  xleftmargin=4pt,
  xrightmargin=4pt,
  aboveskip=5pt,
  belowskip=5pt,
  captionpos=b,
  numbers=none,                        %
  tabsize=2,
  breaklines=false,
  morekeywords={CREATE,TEMP,VIEW,AS,DETECT,SCENES,WHERE,RETURN,TRACK,OBJECTS,SEED,BY,SELECT,FROM,OVER},
}

\lstdefinestyle{ummaPython}{
  language=Python,
  basicstyle=\ttfamily\scriptsize,
  keywordstyle=\bfseries\color{blue!70!black},
  commentstyle=\color{gray},
  stringstyle=\color{teal!60!black},
  showstringspaces=false,
  columns=fullflexible,
  keepspaces=true,
  frame=single,
  rulecolor=\color{black!30},
  frameround=tttt,
  xleftmargin=4pt,
  xrightmargin=4pt,
  aboveskip=5pt,
  belowskip=5pt,
  captionpos=t,
  tabsize=2,
  breaklines=true,
  morekeywords={env,Environment,DetectScenes,VideoFirstPruning,ConditionalLiDAR,DecideFusion,Track,save},
}

\newcommand{\bluecolor}[1]{{{#1}}}

\begin{document}

\title{EMMA: Efficient and Effective Cooperative Multi-Modal Analytics over Embodied AI Data}

\author{Jiangneng Li}
\authornote{Both authors contributed equally to this work.}
\email{jiangnen002@e.ntu.edu.sg}
\affiliation{%
  \institution{Nanyang Technological University}
  \country{Singapore}
}
\author{Jie Wang}
\authornotemark[1]
\email{JWANG128@e.ntu.edu.sg}
\affiliation{%
    \institution{Nanyang Technological University}
  \country{Singapore}
}
\author{Ziqi Yin}
\email{ziqi003@ntu.edu.sg}
\affiliation{%
    \institution{Nanyang Technological University}
  \country{Singapore}
}
\author{Zheng Wang}
\email{zheng011@e.ntu.edu.sg}
\affiliation{%
    \institution{Nanyang Technological University}
  \country{Singapore}
}
\author{Gao Cong}
\email{gaocong@ntu.edu.sg}
\affiliation{%
    \institution{Nanyang Technological University}
  \country{Singapore}
}

\renewcommand{\shortauthors}{Jiangneng Li et al.}

\begin{abstract}

\bluecolor{Embodied AI has seen growing real-world deployment, particularly in autonomous vehicles and robots. These vehicles and robots continuously capture large collections of multi-modal data, such as video, LiDAR point clouds, and GPS trajectories. Historical collections of these data support model improvement and safety assessment through scene retrieval, object trajectory extraction, and corresponding analysis. Processing such analytical queries is challenging because answering them may require combining information from different modalities and observations captured by different moving vehicles or robots, while extracting that information with deep models at every scene incurs prohibitive cost.

We present EMMA, an operator-oriented query-processing framework for 
embodied AI data.
EMMA compiles analytical queries into composable 
operators and executes them on demand. To support accurate query evaluation,
EMMA performs modality-aware staged predicate evaluation and selectively combines overlapping observations. To reduce query cost, EMMA uses a temporal co-location index to identify joint-analysis candidates, adaptive scene skipping to 
reduce model calls,
and a shared world store to reuse extracted 
information
across operators and queries. 
EMMA trains execution policies offline and deploys them as plug-ins. Based on lightweight observations, these policies determine whether the current scene is analyzed independently or jointly and how many subsequent scenes can be safely skipped, balancing analytical accuracy and execution cost.
Experiments show that EMMA substantially reduces query execution cost while preserving query accuracy.
}
\end{abstract}

\maketitle
\pagestyle{\vldbpagestyle}

\section{Introduction}
\label{sec:emma_application_driven_intro}

Embodied AI~\cite{duan2022survey, zang2025rlinf} has seen growing real-world deployment, particularly in autonomous vehicles and robots.
For instance,
Waymo has provided more than ten million
autonomous rides~\cite{waymo}, while Baidu's Apollo Go has provided more than
22 million cumulative auto rides by 2026~\cite{apollo_go}. Tesla has made
Full Self-Driving available on its vehicles~\cite{tesla_fsd_supervised}, and
Starship Technologies operates fleets of autonomous sidewalk robots for
last-mile delivery~\cite{starship}. As these vehicles or robots operate in the physical
world, they continuously 
capture data such as camera streams, LiDAR point clouds, and GPS signals,
creating increasingly large collections of information-rich
multi-modal data.

The recorded data are collected and stored in a server-side data lake for
downstream analysis. Companies analyze these historical data to improve
embodied AI models. For example, developers fine-tune
Vision-Language-Action (VLA)
models~\cite{zitkovich2023rt,kim2024openvla,zhou2026opendrivevla} on
\emph{corner-case scenes}, which we define as scenes in which a deployed model
performs poorly or produces decisions that may lead to safety-critical
outcomes. A central step is to identify representative scenes and reconstruct
the behavior of the involved objects. Beyond model improvement, the same data
can support safety assessment and operational analysis. We consider two
representative applications:

\noindent\textbf{Corner-case retrieval and trajectory extraction.}
An autonomous driving developer may need to retrieve corner-case scenes in which
objects exhibit a particular combination of appearance, 3D location, and
motion, and then track the involved objects to recover their motion trajectories.
The resulting scenes and trajectories can be inspected to diagnose model
failures or added to a training set for model
improvement~\cite{nvidia_alpamayo}.

\noindent\textbf{Safety analysis.}
For example, an analyst may want to determine how frequently autonomous
vehicles encounter dangerous driving situations, such as near-collision or
sudden-braking events, across different cities. Comparing these frequencies
can help developers adapt driving strategies to city-specific traffic
conditions. Such an analysis can first retrieve the relevant scenes, track
the participating objects, and then compute standard aggregates such as event
count, average duration, or minimum separation.

These applications cannot be supported by running a deep model on a single
scene. They require a data system that can compose scene
retrieval, object tracking, and downstream aggregation over large collections
of recorded multi-modal data. Listing~\ref{lst:application-query-new} shows a
representative composed query for corner-case retrieval and trajectory
extraction.

\begin{lstlisting}[style=umma, caption={Example analytical query.}, captionpos=t,
        label={lst:application-query-new},
        numbers=left
        ]
CREATE TEMP VIEW scenes AS
  DETECT SCENES
  WHERE car_1.color='red' AND car_2.color='green'
    AND near(car_1.pos3d, POINT3D(:x1,:y1,:z1), :r1)
    AND near(car_2.pos3d, POINT3D(:x2,:y2,:z2), :r2)
    AND car_1.speed BETWEEN :v1_min AND :v1_max
    AND car_2.speed BETWEEN :v2_min AND :v2_max
  FROM VEHICLES ('V_A','V_B','V_C')
  OVER [10:00,11:00]
  RETURN scene_id, agent_id, timestamp, obj_id, label;
TRACK OBJECTS
  SEED BY (SELECT scene_id, obj_id FROM scenes WHERE label='car_1')
FROM VEHICLES ('V_A','V_B','V_C')
RETURN TRAJECTORY;
\end{lstlisting}

\begin{example}
\label{ex:application-query-new}
The query first retrieves scenes containing two target objects with specified
colors, 3D regions, and speed ranges (lines 1--10). It then tracks
\texttt{car\_1} from the retrieved scenes and returns its trajectory
(lines 11--14). The same retrieved scenes and trajectories can also feed
standard relational aggregates for fleet-level analysis. Because such
aggregates are inexpensive once object states have been materialized, EMMA
focuses its optimization on retrieval and tracking, the two operators that
incur model inference and dominate execution cost.
\end{example}

Supporting these applications remains difficult with existing vehicle data
platforms and task-specific deep models
\cite{nvidia_alpamayo,HuaweiConnectedCar,waymo}. For example, vehicle data
platforms primarily provide ingestion, storage, and management of
heterogeneous sensor streams. The information required by the example query,
such as object colors, locations, and trajectories, is not directly available
in the raw data; it must be extracted by deep models for tasks such as object
detection~\cite{li2025mast}. Meanwhile, existing methods can combine
observations from
different vehicles for predefined tasks such as object detection
\cite{shi2020pv,ruan2024learning,DBLP:conf/icra/XuXXHLM22}, but
they do not provide an execution engine for composing analytical queries over
historical multi-modal data. This leaves a gap between managing embodied AI
data and processing analytical queries over such data efficiently and
effectively.

\begin{figure}[t]
    \centering
    \includegraphics[width=1\linewidth]{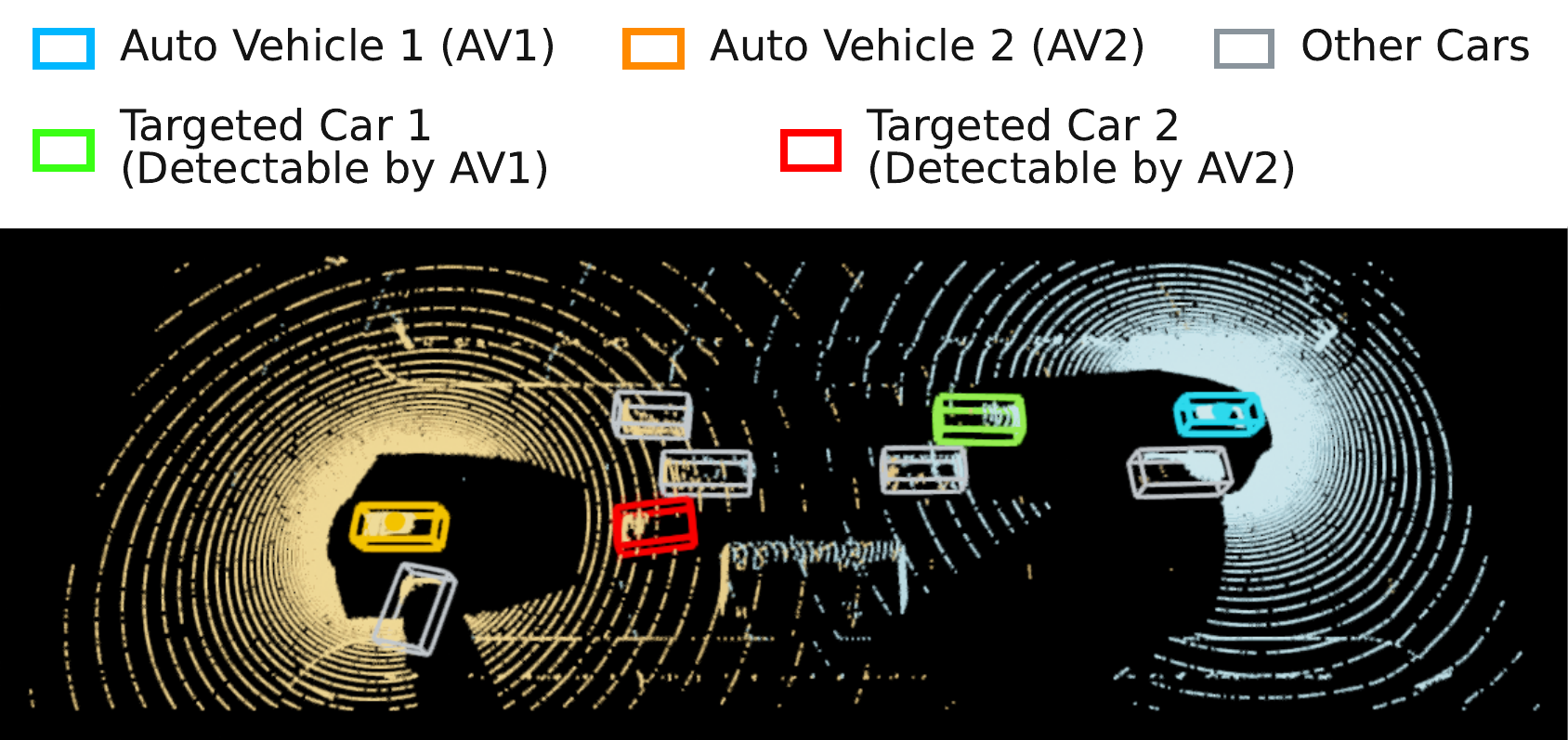}
    \vspace*{-4mm}
    \caption{Complementary vehicle observations for effective scene retrieval.}
    \label{fig:cooperative-observation-new}
    \vspace*{-3mm}
\end{figure}

These analytical queries are difficult to process because they may require information from different modalities and observations recorded by different vehicles or robots, while extracting such information with deep models incurs substantial execution cost. This creates two coupled challenges:

\noindent\textbf{Effective query answering from partial observations.}
Observations recorded by nearby vehicles/robots, such as camera images and
point-cloud frames, can provide complementary views of the same scene. An
object can be occluded, distant, or poorly observed from one vehicle while
remaining visible from another. Figure~\ref{fig:cooperative-observation-new} illustrates
such a case during the scene-retrieval step of
Example~\ref{ex:application-query-new}: the red and green target vehicles/robots are
visible from complementary sensor viewpoints. 
\bluecolor{We term the analysis of data from one vehicle or robot \emph{independent analysis} and the combined analysis of overlapping data from multiple vehicles or robots \emph{joint analysis}. Independent analysis can therefore miss a qualifying scene or prematurely terminate a target trajectory.}
Modalities are complementary as well. In
Listing~\ref{lst:application-query-new}, camera images provide object appearance
information for color predicates, whereas point clouds provide the 3D geometry
needed to evaluate location and motion predicates. GPS-derived trajectories
provide vehicle poses over time for aligning observations across vehicles/robots.
Consequently, no single modality is sufficient to evaluate the complete query.

\bluecolor{However, determining how each scene should be analyzed is non-trivial. Vehicles and robots move independently, so their spatial overlap, viewing directions, and target visibility change over time. Different query predicates may also require information from different modalities. Consequently, a combination of observations and modalities that is sufficient for one scene or query target may be insufficient for another. Whether a candidate combination is useful for the current scene is difficult to determine without actually executing the corresponding deep models. Constructing an effective query- and scene-dependent execution plan using only the limited information available before model execution is therefore challenging.}

\noindent\textbf{Efficient execution of model-intensive queries.}
One naive execution plan is to process every scene with
all applicable models and jointly analyze overlapping observations.
However, this plan is prohibitively expensive over long sensor sequences.
\bluecolor{In the experiments, this exhaustive plan can take more than one day
to process a single query over 200 seconds of data collected by 100 autonomous
vehicles.}
\bluecolor{Much of the query latency comes from repeatedly running expensive models on adjacent scenes with similar content. Joint analysis adds further cost by running models on 
 data from other vehicles or robots
and aligning their outputs.}
\bluecolor{Multiple operators and queries
may 
recompute object states already extracted by earlier operators or queries.
How to reduce query execution cost by limiting model usage and avoiding redundant model execution remains a key challenge.}

\bluecolor{The two challenges are inseparable: processing additional modalities and complementary observations recorded by different vehicles or robots can improve analytical accuracy but also requires more model execution and increases query cost. 
This leads to our %
research question:
\emph{How can a query engine selectively process and combine multi-modal observations from different vehicles or robots to reduce query cost while preserving analytical accuracy?}}

To address this question, we develop \textbf{EMMA}, a system for efficient
and effective cooperative multi-modal analytics over embodied AI data. EMMA
compiles an analytical query into a logical plan composed of operators such as
\texttt{retrieval} and \texttt{tracking}. It executes this plan on demand,
adaptively selecting at each execution step between fast, low-cost analysis and more
accurate but expensive analysis to achieve high query accuracy with low
latency.

To improve effectiveness, 
\bluecolor{EMMA uses lightweight, query- and scene-specific observability features to choose between independent and joint analysis at each retrieval or tracking step.}
For each query filter,
\bluecolor{EMMA performs modality-aware predicate decomposition and staged evaluation: point-cloud analysis evaluates 3D location and motion filters, while image models are invoked only on surviving candidates to verify appearance attributes such as object color.}
EMMA invokes joint analysis only when data from another vehicle or robot
can improve target detection or predicate evaluation, and selects
collaborators dynamically as spatial overlap changes over time.

\bluecolor{To improve efficiency, EMMA reduces unnecessary model execution across candidate observations, modalities, scenes, and queries. 
Specifically, EMMA uses a temporal co-location index to efficiently identify potential joint-analysis candidates based on spatio-temporal overlap. Also, we apply staged predicate evaluation so that additional modality-specific models are invoked only for candidates retained by earlier filters. EMMA also adaptively skips scenes to avoid model calls on every scene in a sequence. Further, EMMA maintains a shared world store that allows later operator steps and subsequent queries to reuse previously extracted object states.}

\bluecolor{Finally, to balance effectiveness and efficiency, EMMA deploys offline-trained, pluggable policies to guide both analysis-mode selection and scene skipping.
Based on lightweight query- and scene-specific state, the policies choose between independent and joint analysis for the current scene and determine how many subsequent scenes can be safely skipped before the next model invocation. This allows EMMA to select more costly analysis when it is likely to improve query accuracy and reduce unnecessary model calls.}

Experiments show that EMMA achieves accuracy competitive with the
best-performing baseline while running over $160\times$ faster on DOMV retrieval.
We summarize our contributions as follows:

\begin{itemize}[leftmargin=10.2pt, itemsep=0pt, parsep=0pt, topsep=0pt]
    \item 
    We develop EMMA, an operator-oriented query-processing framework for multi-modal data recorded by vehicles and robots. EMMA supports composable operators, including scene retrieval and object tracking.
    
    \item We introduce query-conditioned execution over 
    overlapping observations from vehicles and robots.
    EMMA selectively uses independent or joint analysis and
    modality-specific evidence to improve 
    query
    accuracy
    under partial observations.

    \item 
    We design a policy-guided optimizer that reduces model calls by deciding, for each processed scene, whether to use independent or joint analysis and how many subsequent scenes can be safely skipped. A co-location index limits the vehicles or robots considered for joint analysis, while a shared world store allows query operators to reuse previously extracted object states.
    \item 
    We conduct extensive experiments on multiple datasets. The results demonstrate that EMMA improves the accuracy-efficiency trade-off for query processing.
\end{itemize}

\section{Preliminaries} \label{sec:multi-vehicle-preliminary}

\noindent
\bluecolor{\textbf{Multi-Modal Observations.}
Historical multi-modal observations recorded by embodied AI systems,
such as autonomous vehicles and robots~\cite{duan2022survey}, are
collected and stored in a server-side data lake. 
We use $a \in \mathcal{A}$ to denote a vehicle or robot contributing
observations.
In this paper, we consider three modalities: video streams captured
by cameras, point-cloud streams generated by LiDAR sensors, and
locations recorded by GPS modules. Each video frame, point-cloud
frame, and GPS location is associated with a timestamp.
We next describe the representation of observations
in each modality.}

\noindent
\emph{\underline{(1) Video Stream}.}
At timestamp $t$, the camera of 
$a$ produces a 2D image frame observation
denoted by
$\mathit{Img}_{a,t} \in \mathbb{R}^{x \times y \times 3}$,
where $x \times y$ is the fixed spatial resolution and
$\mathit{Img}_{a,t}[u,v] \in \mathbb{R}^{3}$ denotes the RGB vector
of pixel $(u,v)$.
We define the video stream recorded by $a$ as
$
\mathcal{V}_a =
\left\langle
(\mathit{Img}_{a,t_1},t_1),\ldots,
(\mathit{Img}_{a,t_m},t_m)
\right\rangle,
$
where $\mathit{Img}_{a,t_i}$ is the image frame recorded at timestamp
$t_i$. The frames are generated at a frequency of $f_v$ frames per
second.

\noindent
\emph{\underline{(2) Point Cloud Stream}.}
We define the point cloud (PC) frame and PC stream following \cite{li2025mast}. Specifically, a PC frame is defined as a set of multidimensional points: ${\tt P} = \{p_1, \ldots, p_n\}$. Each point $p_i = (x,y,z, F)$, where $(x,y,z)$ denotes a 3D spatial coordinate and $F$ is the additional feature associated with the point.

We then define the PC stream generated by a vehicle/robot as $\mathcal{P} = \langle ({\tt P}_1, t_1, pose_1), \ldots, ({\tt P}_m, t_m, pose_m)\rangle$, where ${\tt P}_i$ is the $i$-th PC frame, $t_i$ is its timestamp, and $pose_i$ is the transformation matrix from the relative coordinate system (centered at the LiDAR sensor) to the global coordinate system. Finally, multiple PC streams generated by different vehicles/robots $\{V_1, \ldots, V_k\}$ are stored in a data lake for downstream analysis, denoted as $\mathcal{D} = \{\mathcal{P}_{V_1}, \ldots, \mathcal{P}_{V_k}\}$. 
The point cloud (PC) frames are generated at a frequency of $f_p$ frames per second. 
In this paper, we assume that
the video and point cloud streams are synchronized, i.e., $f_v = f_p$.

\noindent
\emph{\underline{(3) Trajectory}.}
At timestamp $t$, the GPS module of a vehicle/robot $a$ records a 3D location
$
\mathit{loc}_{a,t}
=
(x_{a,t},y_{a,t},z_{a,t})
\in \mathbb{R}^{3}.
$
We define the trajectory of $a$ as the temporally ordered sequence
$
\mathcal{L}_a =
\bigl\langle
(\mathit{loc}_{a,t_1},t_1),\ldots, 
(\mathit{loc}_{a,t_m},t_m)
\bigr\rangle.
$

\noindent
\textbf{Scene.}
We define a \emph{scene} $s$ as a multimodal snapshot observed by 
$a \in \mathcal{A}$ at a specific timestamp $t$.
Formally, a scene is represented as
\[
s = \left(a, t, \mathit{Img}_t, P_t, \mathit{loc}_t\right),
\]
where $\mathit{Img}_t$ and $P_t$ denote the video frame and point cloud frame captured at time $t$ by $a$,
and $\mathit{loc}_t$ is the 3D location of $a$ obtained from the GPS sensor.
For each $a$, we define its scene stream as
$\mathcal{S}_a=\langle s_{a,t_1},\ldots,s_{a,t_m}\rangle$,
where each scene $s_{a,t_i}$ jointly represents the video frame, point cloud frame,
and location observed by $a$ at timestamp $t_i$.
Given a set of vehicles/robots $\mathcal{A}$, the complete 
scene collection is
denoted as $\mathcal{S}=\{\mathcal{S}_a \mid a\in\mathcal{A}\}$.

\section{Framework Overview}
\label{sec:emma_framework}

\begin{figure}[ht]
    \centering
    \vspace*{-2mm}
    \includegraphics[width=0.82\linewidth]{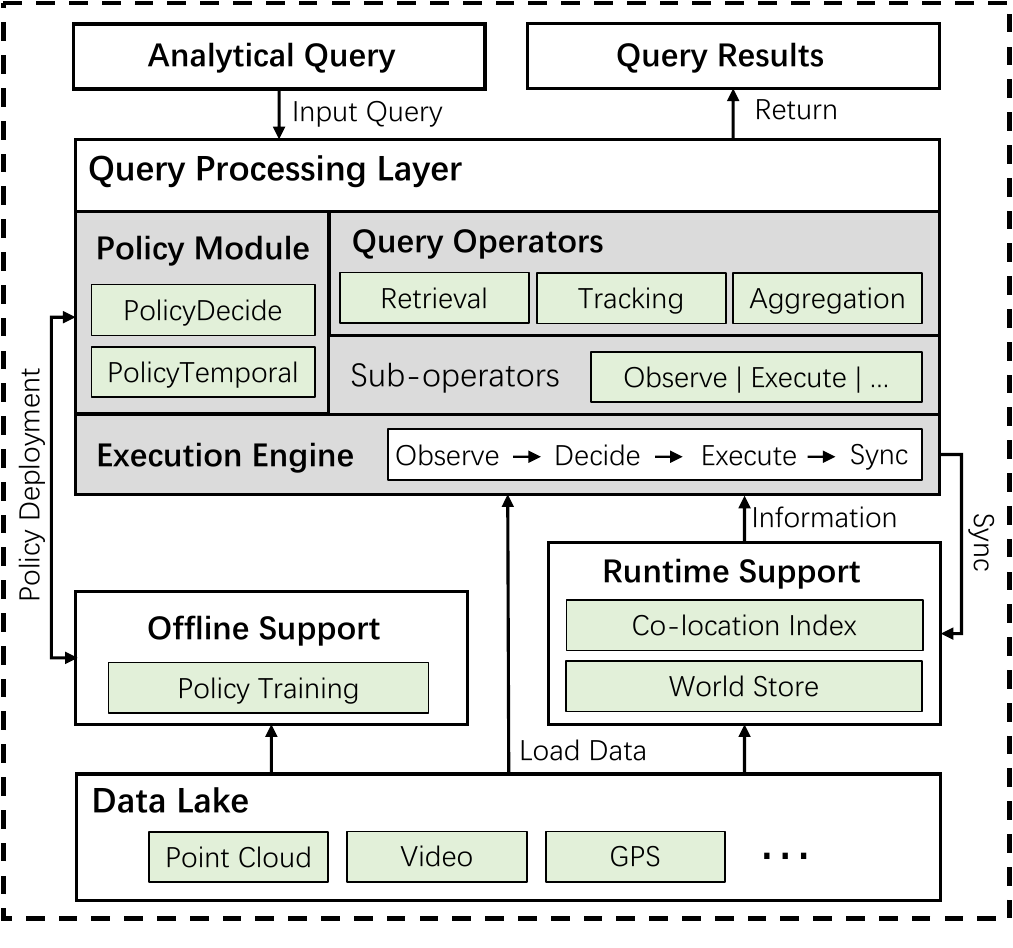}
    \vspace*{-3mm}
    \caption{Framework Overview of EMMA.}
    \vspace*{-3mm}
    \label{fig:framework_overview}
\end{figure}

\noindent
\textbf{Components.}
Figure~\ref{fig:framework_overview} shows the framework of EMMA.
(1) The data lake stores 
multi-modal 
data, including LiDAR, video, GPS, etc.
(2) 
\bluecolor{The offline support module trains execution policies for query
processing. 
At runtime, the query planner invokes these policies to determine whether to perform independent or joint analysis and how many subsequent scenes can be safely skipped.}
(3) \bluecolor{The runtime support module maintains a temporal co-location index that efficiently identifies nearby vehicles or robots with recorded observations for joint analysis, and a world store that caches previously extracted object states and execution information for reuse.}
\bluecolor{(4) The query processing layer contains query operators, 
reusable sub-operators,
a policy module, and an execution engine. 
Query operators specify the analytical tasks.
The retrieval operator identifies scenes that satisfy the query predicates, the tracking operator 
reconstructs the trajectory of a targeted object,
and the aggregation operator computes summary statistics.
Each operator is implemented by composing reusable sub-operators for tasks such as 
 constructing lightweight observations, executing deep models, 
and synchronizing newly extracted information with the world store.
The policy module 
applies the policies to decide the specific execution strategies during query processing.}
The execution engine orchestrates these sub-operators and policies to execute each operator over the scene collections.

\noindent
\textbf{Pipeline.}
\bluecolor{Given 
the example query in Listing~\ref{lst:application-query-new},}
\bluecolor{EMMA compiles the example query into a logical plan consisting of a retrieval operator followed by a tracking operator.}
The retrieval operator searches the selected 
scene collection
within the specified time window and returns scenes containing objects that satisfy the color, location, and speed predicates. The tracking operator then uses the retrieved instances of \texttt{car\_1} as seeds and reconstructs their trajectories.

\bluecolor{The execution engine processes each operator following a specific workflow:
At each timestamp selected for processing,
EMMA first constructs a lightweight observation from the current query state, scene context, and reusable information in the world store ({\tt Observe}). It then invokes \textsc{PolicyDecide} to select the perception source and determine whether to perform independent or joint analysis ({\tt Decide}). Based on this decision, the engine loads the required sensor data, executes the selected perception (model processing) pipeline ({\tt Execute}), and either evaluates the retrieval predicates or updates the tracked target state. Finally, \textsc{PolicyTemporal} determines the next timestamp to process, and the newly extracted object states are synchronized with the world store ({\tt Sync}). 
EMMA passes the scenes returned by the retrieval operator to the tracking operator and returns the extracted trajectories as query results.}

\section{EMMA: 
Cooperative Multi-Modal Analytics}
\label{sec:emma_method}

\bluecolor{We now
introduce the methodology, including indexing, query operators, execution
strategies, and policy-guided optimization.}

\subsection{Indexing}

\noindent
\textbf{Temporal Co-Location Indexing.}
EMMA maintains a global temporal co-location index shared by all 
$a \in \mathcal{A}$.
For each timestamp $t$, the index stores an array of the global positions of all available 
vehicles/robots:
$
\mathcal{I}_t=\{(a,\mathbf{p}_{a,t})\mid a\in\mathcal{A}_t\},
$
where $\mathcal{A}_t$ denotes vehicles/robots available at time $t$ and $\mathbf{p}_{a,t}$ is the global position of $a$.
Given an input location $\ell$, such as the location of a target object,
the index computes distances from $\ell$ to all $a \in \mathcal{A}_t$ in parallel and returns the top-$k$ nearest vehicles/robots:
\[
\mathcal{C}_k(\ell,t)=\mathrm{TopK}_k\left(\{(a,\|\mathbf{p}_{a,t}-\ell\|_2)\mid a\in\mathcal{A}_t\}\right).
\]
\bluecolor{The returned $\mathcal{C}_k(\ell,t)$ 
 forms the
candidate set for joint analysis.
When 
$|\mathcal{A}_t|$ scales up, EMMA can build a lightweight KD-tree for each timestamp $t$ over $\{\mathbf{p}_{a,t}\}$ to reduce the candidate set.}

\subsection{Operators}
\label{sec:emma_operator_def}

Given the 
scene collection $\mathcal{S}$ defined in Section~\ref{sec:multi-vehicle-preliminary}, EMMA supports analytical queries $q$ over $\mathcal{S}$. 
EMMA compiles each query into a logical operator plan, where each operator consumes $\mathcal{S}$ or intermediate results produced by preceding operators, and returns scene sets, trajectories, 
or aggregate statistics.
We next define the core operators supported by EMMA.

\noindent
\textbf{Retrieval.} 
The retrieval operator intends to find 
scenes satisfying the predicates.
Specifically, the retrieval operator 
can
process
over the entire dataset,
or a portion of the dataset 
with a time window.
We formally define the operator as follows: 

\begin{definition}[Retrieval Operator]
    The retrieval operator takes  the predicate filters $F$ 
    (and an optional time window $[t_s,t_e]$, where $t_s$ and $t_e$ denote the start and end 
    times of the window,
    respectively) as input, and returns scenes 
    satisfying $F$. 
\end{definition}

In physical execution, EMMA returns an approximate result due to
deep-model inference and scene skipping.
EMMA supports multiple types of predicates to enable expressive and complex retrieval.
Specifically, we consider the following predicate categories.
(1) 
The \emph{Feature Predicate} constrains object-level attributes, such as object type, color, and shape.
(2) 
The \emph{Spatial Predicate} specifies spatial relationships among objects.
In this paper, we study the
\emph{location-motion filter} that restricts objects to appear within a predefined spatial region,
formalized as
$\bigl\{\{ {\tt obj_1 :} (loc_1, v_1), \ldots, {\tt obj_m} : (loc_m, v_m) \}, \theta_l, \theta_v \bigr\}$,
where $loc_i$ and $v_i$ denote the location and velocity of object $\tt obj_i$,
and $\theta_l$ and $\theta_v$ represent the tolerated error thresholds for location and velocity, respectively. 
Other spatial filters, such as distance constraints, can also be incorporated.

\noindent
\textbf{Tracking.}
The tracking operator identifies and follows a target object, 
and returns its 3D trajectory, which can be further used for downstream analytics.
We formally define the operator as follows.

\begin{definition}[Tracking Operator]
The tracking operator takes as input 
an initial target object specified by a bounding box $b_0$ observed in scene $s_{a,t_0}$,  and an optional 
time window
$\mathcal{T}=[t_0,t_e]$.
It outputs a target trajectory $\hat{\tau}=\{(t,\hat{\mathbf{b}}_t)\mid t\in\mathcal{T}\}$, where $\hat{\mathbf{b}}_t$ denotes the recovered bounding box of the same target at timestamp $t$.
\end{definition}
The operator maintains the target identity across 
time
while maximizing trajectory 
completeness
and localization accuracy.

\noindent
\textbf{Aggregation.} 
\bluecolor{The aggregation operator applies standard functions, such as
\textsc{Count}, \textsc{Avg}, \textsc{Min}, and \textsc{Max}, to the
materialized outputs of retrieval and tracking operators and returns
the aggregate statistics.}

\subsection{Model-Based Query Execution}

Raw multi-modal observations do not directly provide the object-level
information required by
analytical queries.
EMMA therefore invokes
deep models~\cite{DBLP:conf/cvpr/ZhouT18} to extract information such as object bounding boxes and
visual attributes; we refer to this model execution as \emph{perception}.
Because these model invocations are a major source of query cost,
applying them to every scene is expensive.

EMMA reduces model usage along two dimensions. For each processed
scene, it determines whether one observation stream is sufficient or
whether overlapping streams should be analyzed jointly. Independent
analysis is cheaper, whereas joint analysis can recover objects missed
from one viewpoint at additional model and alignment cost. 
Further,
EMMA exploits redundancy among neighboring scenes to avoid invoking
models at every timestamp.

\subsection{Query Processing}
\label{sec:emma_method_query_processing}

\begin{figure*}[t]
    \centering
    \begin{tikzpicture}[
        font=\scriptsize,
        box/.style={
            draw=black!45,
            rounded corners=2pt,
            align=center,
            minimum height=0.62cm,
            text width=1.85cm,
            fill=gray!5
        },
        policy/.style={
            box,
            fill=blue!7,
            draw=blue!45!black
        },
        store/.style={
            draw=black!45,
            rounded corners=2pt,
            align=center,
            text width=6.4cm,
            minimum height=0.62cm,
            fill=orange!8
        },
        arrow/.style={->, thick, draw=black!65},
        hint/.style={->, thick, draw=orange!75!black, dashed}
    ]
    \node[box] (state) at (0,0) {Operator\\State};
    \node[box] (observe) at (2.25,0) {\textsc{Observe}\\build context};
    \node[policy] (decide) at (4.65,0) {\textsc{PolicyDecide}\\select action};
    \node[box] (execute) at (7.05,0) {\textsc{Execute}\\model inference};
    \node[box] (eval) at (9.45,0) {\textsc{Eval/Track}\\evaluate / update};
    \node[policy] (temporal) at (11.85,0) {\textsc{PolicyTemporal}\\skip decision};
    \node[box] (sync) at (14.15,0) {\textsc{SyncWorld}\\store states};
    \node[store] (world) at (7.05,-0.88) {
        \textbf{Shared World Store:}\\[-1pt]
        Object states: boxes, locations, velocities, trajectories
    };

    \draw[arrow] (state.east) -- (observe.west);
    \draw[arrow] (observe.east) -- (decide.west);
    \draw[arrow] (decide.east) -- (execute.west);
    \draw[arrow] (execute.east) -- (eval.west);
    \draw[arrow] (eval.east) -- (temporal.west);
    \draw[arrow] (temporal.east) -- (sync.west);
    \draw[arrow] (sync.north) -- ++(0,0.285) -| (state.north);

    \draw[hint] (world.west) -| node[pos=0.72, left, font=\tiny] {stored states} (observe.south);
    \draw[hint] (sync.south) |- node[pos=0.72, above, font=\tiny] {extracted states} (world.east);
    \end{tikzpicture}
    \vspace*{-2mm}
    \caption{Operator execution pipeline with policy-guided spatial and temporal decisions.}
    \label{fig:execution_pipeline_sketch}
    \vspace*{-3mm}
\end{figure*}
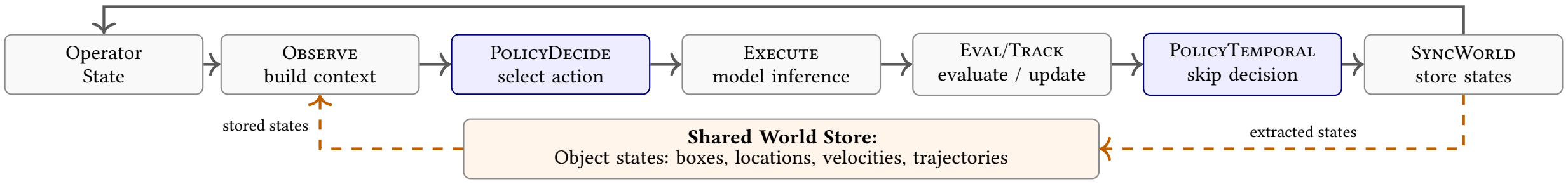

Fig.~\ref{fig:execution_pipeline_sketch} illustrates EMMA's runtime loop for executing an operator. EMMA repeatedly constructs a lightweight observation, selects the execution pipeline with \textsc{PolicyDecide}, executes the chosen pipeline, selects the temporal transition (whether to skip scenes) with \textsc{PolicyTemporal}, and synchronizes extracted evidence into the shared world store for reuse. \bluecolor{We proceed to introduce the sub-operators and  how they form the operators.}

\noindent
\textbf{Sub-Operators.}
To support the main functionalities, we develop multiple sub-operators, including sub-operators that 
apply
deep model to percept the data, and apply specific policy to  decide the query plan.
Alg.~\ref{alg:policy_library} in the appendix gives the pseudocode for these sub-operators.

\smallskip
\noindent\textit{\textsc{Observe}.}
This sub-operator
retrieves a task-aware observation at time $t$ from the scene collection $S$ and augments it with 
information from the world store $\mathcal{W}$.
The returned observation $\texttt{obs}$ serves as a lightweight summary of the current scene context.
\bluecolor{For tracking, \textsc{Observe} applies a Kalman filter~\cite{weng2020ab3dmot} to the 
partial trajectory of the target
recovered at earlier timestamps to predict its bounding box and
velocity at timestamp $t$.}
Expensive 
deep
models are not invoked in \textsc{Observe}.

\smallskip
\noindent\textit{\textsc{PolicyDecide}.}
It
maps the task-specific execution state, including the task type (e.g., retrieval or tracking), predicate,
and current observation, 
into an analysis action $\mathbf{a}$.
\bluecolor{The policy 
chooses
between independent and joint analysis,
and determines whether an additional modality is needed for predicate evaluation.}
For example, in a retrieval operator with spatial and color predicates, the action  first selects 
independent or joint analysis
to obtain candidate boxes satisfying the spatial predicate.
If a candidate passes the spatial filter, the execution pipeline then invokes an 
image 
model to verify the color predicate.
\bluecolor{In a tracking operator, the action selects 
between independent and joint analysis to recover the trajectory 
at the current timestamp.}

\smallskip
\noindent\textit{\textsc{Execute}.}
Function \textsc{Execute} executes the 
execution
pipeline specified by action $\mathbf{a}$ using both the current observation $\texttt{obs}$ and raw sensor data from the scene collection $S$ at time $t$.
It outputs extracted features $\texttt{feat}$ 
required by 
the query,
such as object detections.

\smallskip
\noindent\textit{\textsc{Track}.}
Function \textsc{Track} updates the target state $b$ 
using
the current observation $\texttt{obs}$ and extracted tracking features $\texttt{feat}$ through \textsc{Execute},
and returns a refined target object bounding box $b'$ computed by the KF-based method.

\smallskip
\noindent\textit{\textsc{Eval}.}
Function \textsc{Eval} evaluates whether the extracted features $\texttt{feat}$ satisfy the retrieval predicate $\varphi$, and returns a boolean decision indicating predicate satisfaction. 
The evaluation first performs box-level 
matching between detected bounding boxes and the predicate targets: 
\bluecolor{Each target specified by the predicate must be matched to a distinct
detected box whose 3D center distance from the target location does
not exceed the distance threshold $\theta_l$.}
For a matched object with a velocity predicate, EMMA detects it
in adjacent scenes and uses a Kalman filter to estimate its velocity.
\bluecolor{If $\varphi$ contains visual predicates such as color or shape, EMMA
invokes the visual model after box-level spatial matching
succeeds. \textsc{Eval} returns true only if both the spatial and visual
predicates are satisfied.}

\smallskip

\noindent\textit{\textsc{SyncWorld}.} 
Function \textsc{SyncWorld} synchronizes newly extracted features \bluecolor{(e.g., output of deep models)} into the world store $\mathcal{W}$ 
\bluecolor{under the key $(\textsc{ID}(S),t, \textbf{a})$, where $\textsc{ID}(S)$
identifies the vehicle or robot that recorded the scene, $t$ is the
scene timestamp, and $\textbf{a}$ is the action.}
\bluecolor{This enables EMMA to reuse extracted object states across operator
steps and queries, reducing repeated model invocations.}

\smallskip
\noindent\textit{\textsc{PolicyTemporal}.}
Function \textsc{PolicyTemporal} maps the 
temporal information
$\chi_t$ into a temporal action $\mathbf{u}$ through $\pi_{\text{temp}}$, and then invokes \textsc{TemporalExe} to realize this action.
\bluecolor{Function \textsc{TemporalExe} applies the temporal action selected by
\textsc{PolicyTemporal} and advances execution to timestamp $t'$.
It returns $(\texttt{out}_{\Delta},\texttt{aux}_{\Delta},t')$, where
$\texttt{out}_{\Delta}$ contains additional scenes matched by the retrieval operator when action $u$ triggers fine-grained local search,
or a generated trajectory segment for the tracking operator over the skipped interval $[t, t^\prime]$.
$\texttt{aux}_{\Delta}$ contains the newly extracted object
information.
The task-specific temporal actions
are detailed in Section~\ref{sec:emma_policy_guided_optimization}.}

\bluecolor{Together, these sub-operators provide a common execution interface for retrieval and tracking. They expose the recurring query execution and matching, 
while allowing extracted information to be reused across operators.}

\begin{algorithm}[ht]
\caption{Retrieval Operator Procedure.}
\label{alg:retrieval_op}
\small
\DontPrintSemicolon
\SetKwInOut{Input}{Input}\SetKwInOut{Output}{Output}

\Input{
Vehicle/robot set $\mathcal{A}$, each  $a \in \mathcal{A} $ with a scene stream $\mathcal{S}_a$; \space
Retrieval predicate $\varphi$ with optional time window $\mathcal{T}$; \space
Policy $\pi_{\text{dec}}$, {$\pi_{\text{temp}}$}; world store $\mathcal{W}$ 
}
\Output{Matched scenes $\mathcal{S}_m$ based on $\varphi$}

$\mathcal{S}_m \leftarrow \emptyset$\;
$\mathcal{Q} \leftarrow$ randomized order of $\mathcal{A}$\;

\ForEach{$a \in \mathcal{Q}$}{
  $t \leftarrow \textsc{StartTime}(\mathcal{S}_a,\mathcal{T})$\;
  \While{$t \le \textsc{EndTime}(\mathcal{S}_a,\mathcal{T})$}{
    $\texttt{obs} \leftarrow \textsc{Observe}(\mathcal{S}_a,t,\mathcal{W},\texttt{retrieval})$\;
    $\mathbf{a} \leftarrow \textsc{PolicyDecide}(\pi_{\text{dec}},\texttt{retrieval},\varphi,\texttt{obs})$\;
    $\texttt{feat} \leftarrow \textsc{Execute}(\mathcal{S}_a, t, \mathbf{a},\texttt{obs})$\;
    \lIf{\textsc{Eval}$(\varphi,\mathcal{S}_a,t,\texttt{feat})$}{
      $\mathcal{S}_m \leftarrow \mathcal{S}_m \cup \{\langle a,t\rangle\}$;
    }
    {$\chi_t \leftarrow \langle \mathcal{S}_a,t,\varphi,\mathbf{a},\texttt{obs},\texttt{feat},\mathcal{W}\rangle$}\;
    {\footnotesize{$(\mathcal{S}_{\Delta},\texttt{feat}_{\Delta},t') \leftarrow \textsc{PolicyTemporal}(\pi_{\text{temp}},\texttt{retrieval},\chi_t)$}}\;
    {$\mathcal{S}_m \leftarrow \mathcal{S}_m \cup \mathcal{S}_{\Delta}$}\;
    \ForEach{$\langle u,\texttt{feat}_u\rangle\in\langle t,\texttt{feat}\rangle\cup\texttt{feat}_{\Delta}$}{
  $\mathcal{W}\leftarrow
  \textsc{SyncWorld}(\mathcal{W},\mathcal{S}_a,u,\mathbf{a},\texttt{feat}_u)$\;
}
    {$t \leftarrow t'$}\;

  }
}

\Return $\mathcal{S}_m$\;
\end{algorithm}

\noindent
\textbf{Retrieval Operator Procedure.}
Alg.~\ref{alg:retrieval_op} presents the procedure of the retrieval operator for evaluating a retrieval predicate $\varphi$ over 
$\mathcal{A}$.
The operator processes 
scene streams in a randomized order, and scans each 
scene stream $\mathcal{S}_a$ along the timeline constrained by an optional retrieval window $\mathcal{T}$.
At each time step $t$, it (i) constructs the task-aware observation $\texttt{obs}$ by combining the local scene context with reusable information from world store $\mathcal{W}$; (ii) invokes the policy $\pi_{\text{dec}}$ to decide an analysis action; (iii) executes the selected pipeline to extract predicate-relevant features; and (iv) evaluates whether $\varphi$ 
is satisfied to report matched scenes.
\bluecolor{To enable information reuse,
the operator synchronizes the extracted features back into $\mathcal{W}$.}
 \bluecolor{Then, a temporal policy $\pi_{\text{temp}}$ chooses the next temporal transition: a coarse transition action skips unpromising scenes, while a fine-grained search action invokes the deep model within a local temporal window where matching scenes are likely to appear.}

As shown in Alg.~\ref{alg:retrieval_op},
the input includes 
$\mathcal{A}$, 
scene stream $\mathcal{S}_a$ of each $a \in \mathcal{A}$, predicate $\varphi$, an optional time window $\mathcal{T}$, the  policies $\pi_{\text{dec}}$ and {$\pi_{\text{temp}}$}, and the world store $\mathcal{W}$.
The output  $\mathcal{S}_m$
records the matched scenes satisfying $\varphi$.
We initialize the match set $\mathcal{S}_m$  and generate a  processing order $\mathcal{Q}$ over $a \in \mathcal{A}$.
We then iterate through $a$s in $\mathcal{Q}$  and initialize the scan start time $t$ for each 
scene 
stream
$\mathcal{S}_a$, constrained by the optional window $\mathcal{T}$ ({lines~1--4}).
Within each selected 
scene stream $\mathcal{S}_a$,
\bluecolor{we scan along time until reaching the end of the retrieval window ({line~5}) and construct the 
lightweight
observation $\texttt{obs}$ from the local scene context, which captures visibility and occlusion cues,} and reusable information retrieved from the world store $\mathcal{W}$ ({line~6}).
The policy $\pi_{dec}$ decides an analysis action $\mathbf{a}$ based on $(\varphi,\texttt{obs})$ ({line~7}), and EMMA executes the corresponding execution pipeline to extract predicate-relevant features $\texttt{feat}$ from both $\texttt{obs}$ and raw data in $\mathcal{S}_a$ at time $t$ with deep model ({line~8}).
EMMA evaluates the predicate $\varphi$ on the extracted features and 
\bluecolor{involves visual models to verify additional color/shape feature if needed,}
and records $\langle a,t\rangle$ in $\mathcal{S}_m$ when 
the scene satisfies $\varphi$
({line~9}).
Afterwards,
EMMA packs the temporal-related context $\chi_t$ ({line~10}) and invokes \textsc{PolicyTemporal}.
\bluecolor{If the selected action triggers fine-grained search, EMMA obtains additional matched scenes $\mathcal{S}_{\Delta}$.}
Then, EMMA decides the next timestamp $t'$ to be processed, merges $\mathcal{S}_{\Delta}$ into $\mathcal{S}_m$, synchronizes 
new extracted information back
into the world store $\mathcal{W}$, and updates the 
time to $t'$ ({line~11--14}).
Finally, EMMA returns the matched results $\mathcal{S}_m$.

\begin{algorithm}[ht]
\caption{Tracking Operator Procedure.}
\label{alg:tracking_op}
\small
\DontPrintSemicolon

\SetKwInOut{Input}{input}\SetKwInOut{Output}{output}

\Input{
Anchor scene stream $S$; initial target state $b_0$ at time $t_0$;\;
Tracking identity $\psi$; unified policy $\pi_{\text{dec}}$, {$\pi_{\text{temp}}$}; world store $\mathcal{W}$
}
\Output{Trajectory $\tau$}

$\tau \leftarrow [\ ]$\;
$(b,t) \leftarrow (b_0,t_0)$\;

\While{\textsc{Valid}$(b)$}{

    {$\texttt{obs} \leftarrow \textsc{Observe}(S,t,\mathcal{W},\texttt{tracking})$}\;
    $\mathbf{a} \leftarrow \textsc{PolicyDecide}(\pi_{dec},\texttt{tracking},\psi,\texttt{obs})$\;
    {$\texttt{feat} \leftarrow \textsc{Execute}(S,t,\mathbf{a},\texttt{obs})$}\;
    $b \leftarrow \textsc{Track}(b,\texttt{obs},\texttt{feat})$\;
    $\tau \leftarrow \tau \mathbin{+\!\!+} [\langle t,b\rangle]$\;
    {$\chi_t \leftarrow \langle S,t,\psi,b,\mathbf{a},\texttt{obs},\texttt{feat},\mathcal{W}\rangle$}\;
    {\footnotesize {$(\tau_{\Delta},\langle  \texttt{feat}_{\Delta}, b' \rangle,t') \leftarrow \textsc{PolicyTemporal}(\pi_{\text{temp}},\texttt{tracking},\chi_t)$} }\;
    {$\tau \leftarrow \tau \mathbin{+\!\!+} \tau_{\Delta}$}\;
    \ForEach{$\langle u,\texttt{feat}_u\rangle\in\langle t,\texttt{feat}\rangle\cup\texttt{feat}_{\Delta}$}{
  $\mathcal{W}\leftarrow
  \textsc{SyncWorld}(\mathcal{W},\mathcal{S},u,\mathbf{a},\texttt{feat}_u)$\;
}
    {$(b,t) \leftarrow (b',t')$}\;

}

\Return $\tau$
\end{algorithm}

\noindent
\textbf{Tracking Operator Procedure.}
Alg.~\ref{alg:tracking_op} describes the tracking operator, which follows a target object across time starting from an anchor scene stream.
Given an initial target object's bounding box at time $t_0$, the operator iteratively observes the current scene context, invokes a policy to select an appropriate tracking pipeline, and updates the 
trajectory $\tau$
accordingly.
The extracted 
object information
is synchronized into the world store for reuse.
Similar to retrieval, a temporal policy adaptively determines the next tracking transition \bluecolor{(the next scene requiring model processing)} to improve efficiency while preserving tracking continuity.
\bluecolor{The key safeguard is a rollback mechanism: when a large temporal jump lands on an unreliable tracking state, e.g., when the target is lost, EMMA rolls back to the skipped region and resumes conservative scene-by-scene evaluation to recover target association. We refer to this mechanism as backjump.}

{The pseudocode of tracking is given in Alg.~\ref{alg:tracking_op}.}
The input includes the anchor scene sequence $S$, the initial target state $b_0$ at time $t_0$, the tracking identity $\psi$, the policies $\pi_{\text{dec}}$ and {$\pi_{\text{temp}}$}, and the world store $\mathcal{W}$.
The output is the trajectory $\tau$ of the target object.
\bluecolor{We initialize the trajectory $\tau$ ({line~1}) and current target state $(b,t)$ with $(b_0,t_0)$ ({line~2}), where $b_0$ is the initial bounding box at time $t_0$.}
While the target state remains valid ({line~3}), EMMA constructs the tracking-aware lightweight observation from the anchor stream and world store, including the Kalman-filter predicted target state ({line~4}), invokes policy $\pi$ to select a tracking action $\mathbf{a}$ ({line~5}), executes the selected execution pipeline ({line~6}), and generates the new bounding box $b$
with the extracted features ({line~7}).
EMMA appends the recovered $b$ to $\tau$ ({line~8}).
\bluecolor{After that, EMMA packs the temporal context $\chi_t$ (line~9) and invokes \textsc{PolicyTemporal} to decide the next timestamp $t'$ to process (line~10). If the selected action skips timestamps, EMMA propagates the current target state across the skipped interval, appends the resulting trajectory segment $\tau_{\Delta}$, and generates the new bounding box $b^{\prime}$. }
\bluecolor{The temporal policy selects a safe action, and \textsc{TemporalExe} invokes the backjump fallback 
when the target
cannot be reliably recovered 
after a skip.}
When the selected action skips timestamps, \textsc{TemporalExe} generates $\tau_{\Delta}$ by propagating the target bounding box over the skipped interval.
EMMA appends $\tau_{\Delta}$ to $\tau$ (line~11), synchronizes 
the extracted information
into $\mathcal{W}$ (line~12--13), and updates $(b,t)$ to $(b',t')$ (line~14).
Finally, EMMA returns the completed trajectory $\tau$.

\subsection{Policy-Guided Optimization} \label{sec:emma_policy_guided_optimization}
We now introduce the policy design 
during operator execution.
Specifically, a policy maps the 
scene context
into an execution action.
Following the query processing workflow in Section~\ref{sec:emma_method_query_processing}, EMMA makes two recurring policy decisions.
At each processed scene, \textsc{PolicyDecide} selects a deep model perception action. 
 \textsc{PolicyTemporal} chooses the next temporal transition to process.
Together, these decisions determine where perception is applied in space and time. We then 
introduce the details of the two policies.

\subsubsection{Policy Inputs and Action Spaces}
\begin{table*}[!t]
    \centering
    \caption{Overview of policy supervision in EMMA.}
    \vspace*{-2.5mm}
    \label{tab:policy_training_overview}
    \footnotesize
    \setlength{\tabcolsep}{4pt}
    \renewcommand{\arraystretch}{1}
    \resizebox{\textwidth}{!}{%
    \begin{tabular}{@{}p{0.055\textwidth} p{0.10\textwidth} p{0.28\textwidth} p{0.33\textwidth} p{0.31\textwidth}@{}}
    \hline
    \textbf{Operator} &
    \textbf{Policy} &
    \textbf{Input State} (Section~\ref{sec:emma_env_observation}) &
    \textbf{Oracle Derived Supervision} (Section~\ref{sec:emma_learned_policy}) &
    \textbf{Training Goal / Loss} (Section~\ref{sec:emma_learned_policy}) \\
    \hline \hline
    \multirow{2}{*}{Retrieval} &
    \textsc{PolicyDecide} &
    Observability of predicate-specified target-box regions over candidate perception actions &
    Action labels indicating successful detection of retrieval targets, with candidate-helper rankings
    &
    Select Ind/Joint action and rank helpers; classification with cross entropy~\cite{DBLP:journals/gpem/Heaton18} + ranking loss
    \\
    &
    \textsc{PolicyTemporal} &
    Detection-conditioned matching feature $\mathbf{z}_t$ &
    Utilities of coarse-skip and fine-search temporal actions &
    Estimate utility of skip/search actions; weighted regression
    \\
    \midrule
    \multirow{2}{*}{Tracking} &
    \textsc{PolicyDecide} &
    Observability of the trajectory-predicted target-object box region over candidate perception actions &
    Associable actions ranked by localization quality and collaboration cost &
    Classify the best tracking action; classification with quality and cost loss
    \\
    &
    \textsc{PolicyTemporal} &
    Trajectory-conditioned skip feature $\mathbf{g}_t$ &
    Maximum safe skip length under oracle trajectory recovery &
    Predict maximum safe skip length; classification loss
    \\
    \hline
    \end{tabular}
    }
    \vspace{-1.5mm}
\end{table*}

For a 
query
operator
$o\in\{\texttt{retrieval},\texttt{tracking}\}$, EMMA defines two 
policies:
\[
\pi_{\mathrm{dec}}^o:\mathcal{X}_{\mathrm{dec}}^o\rightarrow\mathcal{A}_{\mathrm{dec}}^o,
\qquad
\pi_{\mathrm{temp}}^o:\mathcal{X}_{\mathrm{temp}}^o\rightarrow\mathcal{A}_{\mathrm{temp}}^o.
\]
$\pi_{\mathrm{dec}}^o$ corresponds to the \textsc{PolicyDecide} and returns a deep model analysis action $\mathbf{a}\in\mathcal{A}_{\mathrm{dec}}^o$. $\pi_{\mathrm{temp}}^o$ corresponds to the  \textsc{PolicyTemporal} and returns a temporal action $\mathbf{u}\in\mathcal{A}_{\mathrm{temp}}^o$.

\smallskip
\noindent
\textbf{\textsc{PolicyDecide}: Analysis Policy.}
\textsc{PolicyDecide} selects the analysis action for the current 
operator state.
Because observability varies across vehicles/robots, we score candidate 
independent and joint analysis
actions instead of relying on a fixed 
$a \in \mathcal{A}$
or a fixed collaboration pair.
An independent action runs the 
deep model 
on one $a \in \mathcal{A}$,
while a 
joint
action runs the 
deep model 
with
observations from two vehicles/robots.

Both action types are 
necessary
because query targets may have different visibility patterns across $a \in \mathcal{A}$.
\bluecolor{We distinguish two cases:}
(1) \textit{Single-view Sufficient.} All query-relevant targets are visible from one $a$, so independent analysis is sufficient.
(2) \textit{Partially Occluded.} One $a_i$ observes only part of the target set, while another $a_j$ provides complementary observations.
Joint analysis actions improve query quality in the 
second case,
but they are more expensive than independent analysis actions.
\textsc{PolicyDecide} therefore chooses between independent and joint analysis actions.

\bluecolor{For retrieval at timestamp $t$, let $e$ denote the vehicle or robot
that recorded the scene currently being evaluated; we refer to $e$ as
the anchor. Let $\mathcal{H}_{e,t}$ denote the candidate vehicles or
robots returned by the co-location index. An independent action analyzes
observations recorded by $e$ alone, whereas a joint action combines
observations recorded by $e$ and a helper $h\in\mathcal{H}_{e,t}$.
The goal is to select the action most likely
to correctly evaluate the query predicates for the current scene. We
define the action space as follows:}
\[
\mathcal{A}_{\mathrm{dec}}^{\mathrm{ret}}(e,t)
=
\{(\texttt{Ind},e)\}
\cup
\{(\texttt{Joint},e,h)\mid h\in\mathcal{H}_{e,t}\},
\]
{where $\texttt{Ind}$ denotes 
independent analysis
using the selected $a$ only, while $\texttt{Joint}$ denotes 
joint analysis
using a vehicle/robot pair.}
{For example, action $(\texttt{Joint},e,h)$ means that the policy selects $e$ and $h$ to jointly analyze the current scene.}

{The selected retrieval action may be followed by a predicate-triggered 
evaluation
such as image/video-based color or shape verification, after box-level matching succeeds.}

For tracking execution, the spatial action set is defined over the candidate vehicle/robot set rather than around a fixed ego-helper context.
\bluecolor{The system first 
predicts the target's bounding box using a KF-based method at the current timestamp,}
and then selects an analysis action between $\texttt{Ind}$ and $\texttt{Joint}$ to update the target trajectory.
\bluecolor{Unlike retrieval, tracking does not evaluate a query predicate at the
current scene. Instead, it selects the action most likely to detect the target near
the position predicted from its trajectory recovered so far, allowing
EMMA to extend the trajectory.}
We formally define the action space with candidate vehicle/robot set $\mathcal{C}_t$ collected from the index structure:
\[
\begin{aligned}
\mathcal{A}_{\mathrm{dec}}^{\mathrm{trk}}(t)
=&\ \{(\texttt{Ind},c)\mid c\in\mathcal{C}_t\}\\
&\cup
\{(\texttt{Joint},c_i,c_j)\mid
\substack{c_i,c_j\in\mathcal{C}_t\\ c_i\neq c_j}\},
\end{aligned}
\]
where each pair $(c_i,c_j)$ represents 
\bluecolor{a candidate joint-analysis
action 
to detect the target near its
predicted position.}

\smallskip
\noindent
\textbf{\textsc{PolicyTemporal}: Temporal Policy.}  \label{sec:emma_skip}
After \textsc{PolicyDecide} selects an analysis action, EMMA executes the action and updates the query state with the resulting 
model outputs.
Because 
deep model calls are costly, \textsc{PolicyTemporal} decides when the next model call should be made. For retrieval execution,  it chooses between coarse scene skipping and fine local search (i.e., searching the nearby temporal window scene by scene); for tracking execution, it predicts the maximum safe skip length.

For retrieval execution, we define two types of  actions: (1) the coarse action: $\{(\texttt{Coarse},s)\mid s\in\{4,8,12\}\}$. Coarse action skips forward by $s$
time steps,
which reduces cost but may miss targeted scenes.
(2) The fine action: $\{(\texttt{Fine},w)\mid w\in\{1,2,3\}\}$.
\texttt{Fine} runs deep model detections 
over a temporal window of $w$ scenes to search for target scenes at fine granularity.
The fine action performs deep model perception 
with the same analysis method selected by \textsc{PolicyDecide} (\texttt{Ind} or \texttt{Joint}).
We define the action spaces as:
\[
\mathcal{A}_{\mathrm{temp}}^{\mathrm{ret}}
=
\{(\texttt{Coarse},s)\mid s\in\{4,8,12\}\}
\cup
\{(\texttt{Fine},w)\mid w\in\{1,2,3\}\}.
\]
\bluecolor{To accommodate different query-cost budgets, EMMA uses a
scene-skipping ratio $r_s$ to scale each coarse action: an action with
base skip length $s$ skips $\lfloor r_s\cdot s\rfloor$ scenes.}

For tracking execution, a temporal action specifies how many intermediate frames can be safely skipped after the current model frame.
{Given a skip action $k$, the system 
predicts the trajectory of the targeted object using a Kalman filter(KF)~\cite{weng2020ab3dmot} based method,
and then performs a spatial perception step at the $k$-th frame to correct the non-model predicted trajectory.
We select $k$ from $\{0,1,3,5,7,10\}$, where $k=0$ means no skip.
Larger values of $k$ reduce model calls but increase the risk of trajectory breaks when motion prediction drifts, occlusion changes, or the target cannot be re-associated at the landing frame.
Specifically, \textsc{TemporalExe} realizes the skip through KF method and applies backjump-aware fallback when the prediction is not reliable.
We formally define the action space as:
\[
\mathcal{A}_{\mathrm{temp}}^{\mathrm{trk}}
=
\{k\mid k\in\{0,1,3,5,7,10\}\},
\]

\subsubsection{Environmental Observation}  \label{sec:emma_env_observation}
EMMA uses lightweight observations as policy input.
These observations provide lightweight features that characterize the current execution context, including the physical environment, 
candidate collaborators,
and historical execution context. The policies use these features to score candidate actions without invoking deep models. 
{Detailed feature computation is provided in Appendix~\ref{app:policy_observation_details}.}

\smallskip
\noindent
\textbf{\textsc{PolicyDecide}: Conditioned Observability Encoding.}
\textsc{PolicyDecide} uses target-conditioned observability features to compare candidate $a \in \mathcal{A}$ and vehicle/robot pairs.
For each target box under a candidate $a$'s viewpoint,
EMMA computes a compact feature $\mathbf{f}=[v,o,p,g,r,\cos\phi,\sin\phi]$, 
\bluecolor{where the first four terms summarize LiDAR-based visibility and
occlusion, while the last three encode
the target's distance from candidate $a$
and its angle relative to $a$'s forward direction.}
For retrieval execution, the current query predicate may specify multiple target objects.
EMMA computes one observability feature list for each predicate-specified target object over the ego vehicle/robot and helper candidates.
For tracking execution, EMMA first predicts the current target box from the previous tracking state, and then computes the same observability feature for each candidate in $\mathcal{C}_t$.
These features let \textsc{PolicyDecide} estimate whether a single $a \in \mathcal{C}_t$ is sufficient or whether another viewpoint may provide useful complementary evidence.

\smallskip
\noindent
\textbf{\textsc{PolicyTemporal}: Temporal State Feature Encoding.}
\textsc{PolicyTemporal} uses temporal state features to summarize whether execution should continue conservatively or skip ahead.
\bluecolor{For retrieval, EMMA encodes the temporal state as a compact vector
$\mathbf{z}_t$ that summarizes detection-to-target distance statistics ($d_{\min}$, $d_{\mathrm{2nd}}$, $d_{\mathrm{mean}}$, $d_{\max}$, $d_{\mathrm{gap}}$),
strict/near object match counts and ratios ($n_{\mathrm{strict}}$, $n_{\mathrm{near}}$, $r_{\mathrm{strict}}$, $r_{\mathrm{near}}$), and the number of query targets ($N_q$):}
\[
\mathbf{z}_t =
[
d_{\min},
d_{\mathrm{2nd}},
d_{\mathrm{mean}},
d_{\max},
d_{\mathrm{gap}},
n_{\mathrm{strict}},
n_{\mathrm{near}},
r_{\mathrm{strict}},
r_{\mathrm{near}},
N_q
].
\]
{For tracking execution, EMMA represents the temporal state with a trajectory-conditioned skip feature $\mathbf{g}_t$.}
{It includes recent velocity history ($v_1,v_2,v_3,v_4$) of the target, acceleration history ($a_1,a_2$), predicted target visibility and occlusion scores ($s_{\mathrm{vis}}, o_{\mathrm{vis}}$), and recent backjump behavior ($r_{\mathrm{back}}$):
$
\mathbf{g}_t =
[
v_1,v_2,v_3,v_4,
a_1,a_2,
s_{\mathrm{vis}},
o_{\mathrm{vis}},
r_{\mathrm{back}}
].
$

{These temporal observations capture whether nearby scenes are likely to contain value scenes for retrieval, or whether the tracking state can be safely propagated without a detector call.}

\subsubsection{Heuristic policies.} 
We provide heuristic policies that decide actions using fixed rules without learning.
For retrieval operator, the \textsc{PolicyDecide} heuristic selects 
\texttt{Ind} action
when the computed 
independent
visibility score of the query targets is $>\theta_v$; otherwise, it selects 
\texttt{Joint} action.
\textsc{PolicyTemporal} 
selects \texttt{Fine} action
if the minimum matching distance $d_{\min}$ between the current model detections and the query targets is $<\theta_d$.
Otherwise, it selects \texttt{Coarse} 
\bluecolor{action and uses the query's
maximum-speed bound to estimate how many scenes can be safely skipped.}
For tracking operator, the heuristic \textsc{PolicyDecide} selects among 
\texttt{Ind} and \texttt{Joint} actions
based on the visibility score and occlusion ratio
of the predicted target.
\textsc{PolicyTemporal} applies a fixed skip length $k_h$ when the recovered trajectory is classified as constant-velocity or constant-acceleration motion, and otherwise 
does not skip any scenes.

\subsubsection{Learning-based Policy} \label{sec:emma_learned_policy}
The heuristic-based rules cannot fully handle query- and context-dependent scenarios, which frequently arise in realistic settings. Therefore, we further 
develop learning-based policies

\smallskip
\noindent
\textbf{Training Overview.}
EMMA trains all policies offline from oracle-guided execution traces.
Table~\ref{tab:policy_training_overview} summarizes the input state, oracle-derived supervision, and training goal/loss for each policy.
Detailed oracle construction, label construction, training procedure, and loss functions are provided in Appendix~\ref{app:policy_training_details}. Specifically, the training pipeline contains four steps.

\smallskip
\noindent
(1) \emph{Oracle reference construction.} EMMA first constructs operator-specific oracle references. For retrieval, the oracle reference is the set of value scenes that satisfy the predicate and are recoverable by at least one 
candidate 
analysis
action. 
For tracking, the oracle reference is the best recoverable target trajectory obtained by exhaustively evaluating all candidate 
analysis
actions over time.

\smallskip
\noindent
(2) \emph{Supervision generation.} EMMA then replays candidate 
actions against these references to generate compact supervision signals. 
\textsc{PolicyDecide}
learns which 
analysis action
should be selected at a processed scene, while 
\textsc{PolicyTemporal}
learns whether execution should move conservatively or skip ahead.

\smallskip
\noindent
(3) \emph{Policy training.} The training order follows the dependency structure of each operator. 
For retrieval, oracle value scenes directly determine temporal labels (i.e., value scenes should not be skipped), so EMMA trains \textsc{PolicyTemporal} before \textsc{PolicyDecide}. For tracking, temporal labels depend on whether the target can be recovered and associated after a skip rollout, so EMMA trains \textsc{PolicyDecide} before \textsc{PolicyTemporal}.

\smallskip
\noindent
(4) \emph{Objective optimization.} Finally, each policy is optimized with a training goal and loss matched to its oracle-derived supervision. Retrieval \textsc{PolicyDecide} selects the Ind/Joint action and ranks helpers, retrieval \textsc{PolicyTemporal} estimates skip/search utilities, tracking \textsc{PolicyDecide} classifies the best tracking action, and tracking \textsc{PolicyTemporal} predicts the maximum safe skip length.

Experimental results in Section~\ref{sec:emma_exp_result} demonstrate the benefit of learned policies over heuristic-policy baselines.

\section{Evaluation}
\label{sec:emma_evaluation}

We aim to evaluate the following research questions:
\textbf{RQ1.} How effective and efficient is EMMA on retrieval workloads, including varying cardinalities and color/shape predicates?
\textbf{RQ2.} How accurately and efficiently does EMMA support object tracking?
\textbf{RQ3.} How effective is EMMA on composed retrieval-and-tracking workloads?
\textbf{RQ4.} How much does shared world-store reuse improve repeated query processing on the same data (Section~\ref{sec:emma_method_query_processing})?
\textbf{RQ5.} What is the runtime cost breakdown of EMMA's major components?
\textbf{RQ6.} How does the retrieval skip range affect the accuracy-efficiency trade-off (Section~\ref{sec:emma_skip})?
\textbf{RQ7.} How do the learning-based policies converge during offline training (Section~\ref{sec:emma_learned_policy})?

\subsection{Experimental Setting}

We proceed to introduce the evaluation setting, including the dataset, the query workload, the comparison methods, and the metrics.

\noindent
\textbf{Dataset.} 
\bluecolor{We use three datasets: OPV2V~\cite{DBLP:conf/icra/XuXXHLM22}, a widely used public benchmark
in which multiple vehicles operate in a shared environment; DOMV,
which models realistic scenarios where vehicles move independently
and their spatial overlap changes dynamically; and DOMV-Long, a
larger-scale extension of DOMV with more vehicles and longer scene
streams.}

\begin{itemize}[leftmargin=10.2pt, itemsep=0pt, parsep=0pt, topsep=0pt]
    \item \textbf{OPV2V}~\cite{DBLP:conf/icra/XuXXHLM22}.  
    It contains 73 short driving
scene streams. 
At each time step, observations are recorded from 2 to 7 connected vehicles and have relatively stable spatio-temporal overlap.
    \item \textbf{DOMV}. \bluecolor{We developed the \underline{D}ynamic \underline{O}verlap among \underline{M}ultiple \underline{V}ehicles (DOMV) Dataset. Specifically, we construct DOMV with the CARLA system~\cite{DBLP:conf/corl/DosovitskiyRCLK17} to evaluate EMMA in 
    scenes where sensor-equipped vehicles move independently and the
overlap between their observed areas changes over time.}
    Unlike 
    OPV2V,
    DOMV features 30 vehicles (10 sensor-equipped smart vehicles and 20 background vehicles) moving with independent trajectories and varying relative velocities. Each smart vehicle records 1,000 continuous scenes of synchronized video, point clouds, and GPS location. 
    It has a total size of 30 GB.
    We ensure vehicles continuously have neighboring vehicles with sensors and evaluate whether EMMA can handle this scenario.
    \item \textbf{DOMV-Long}. 
    It is a larger scale dataset with 300 vehicles \bluecolor{(100 vehicles equipped with sensors)}. 
    Each sensor-equipped vehicle records 2000 continuous scenes. 
    It has a total size of 600 GB.
\end{itemize}

\noindent
\textbf{Query Workload.} 
We designed a comprehensive query workload focusing on the two core analytical operators: retrieval and tracking. 
\bluecolor{In our prototype, standard aggregation is applied directly to the
materialized outputs of other operators, such as retrieval and tracking,
and does not invoke deep models; we therefore do not evaluate it as a
separate workload.}

\begin{itemize}[leftmargin=10.2pt, itemsep=0pt, parsep=0pt, topsep=0pt]
    \item \textbf{Retrieval Workload}.  To evaluate retrieval performance, we sample scenes and target objects to construct predicates with varying complexity and cardinality. Specifically, we vary the query cardinality by adjusting the number of target objects and apply diverse constraints on 3D locations, target velocities, and tolerance thresholds ($\theta_{L}, \theta_{D}$). Additionally, we incorporate multi-modal feature predicates, such as object color,
    to generate a comprehensive workload with diverse selectivity and attribute-filtering requirements.
    \item \textbf{Tracking Workload}. The tracking workload 
    is
    designed to test the system's ability to reconstruct trajectories.
    It contains a robust set of targeted ground-truth trajectories with varying lengths
    to evaluate tracking accuracy and localization precision.
    \item \textbf{Composed Workload}. 
    The composed workload consists of a retrieval operator followed by a tracking operator that tracks the trajectory of the retrieved target object.
\end{itemize}

\noindent
\textbf{Comparison Methods.} 
We select baselines representing three ways of processing each scene:
independently analyzing data from one vehicle (PointPillar~\cite{DBLP:conf/cvpr/LangVCZYB19}), jointly analyzing data
from it and its nearest neighbor (VoxelNet-Near~\cite{DBLP:conf/cvpr/ZhouT18}), and jointly analyzing data from all
available vehicles (VoxelNet-All).
We also
include MAST, a representative point-cloud retrieval system, and
EMMA-H, a heuristic-policy variant of EMMA.
For
consistency, all methods share data loading and preprocessing,
isolating differences in
execution strategy.

\begin{itemize}[leftmargin=10.2pt, itemsep=0pt, parsep=0pt, topsep=0pt]
    \item \textbf{PointPillar}~\cite{DBLP:conf/cvpr/LangVCZYB19}. It
    performs single-vehicle \emph{independent analysis} when extracting semantic information from point clouds.
    \item \textbf{VoxelNet-Near}~\cite{DBLP:conf/cvpr/ZhouT18}. 
    It performs multi-vehicle \emph{joint analysis} using the nearest cooperative vehicle as the auxiliary source.
    \item \textbf{VoxelNet-All}~\cite{DBLP:conf/cvpr/ZhouT18}. 
    It performs exhaustive multi-vehicle joint analysis over all available cooperative vehicles. It serves as a strong collaborative baseline with high computation cost.
    \item \textbf{MAST}~\cite{li2025mast}. 
    It supports location-based retrieval 
    on point cloud data.
    We adapt 
    it 
    to the
    EMMA system, and compare its performance on the retrieval workload.
    \item \textbf{EMMA-H}. Our system with heuristic-based policies. We define explicit heuristic rules for both policies.
    \item \textbf{EMMA}. Our system with learned policies. 
    We train the policies on one half of DOMV and evaluate them on the other half, and apply them to other datasets without fine-tuning to assess cross-dataset generalizability.
\end{itemize}
Specifically, the 
deep models
are pretrained on the OPV2V benchmark~\cite{DBLP:conf/icra/XuXXHLM22} and available in the OpenCood Repository~\cite{opencood}.

\noindent
\textbf{Metrics.}
We evaluate the following metrics:
(1) \emph{Retrieval Accuracy}: \textbf{F1 Score}. Following~\cite{li2025mast}, we apply $F1 = 2 \cdot (\text{Prec} \cdot \text{Rec})/(\text{Prec} + \text{Rec})$, where $\text{Prec} = \frac{TP}{TP + FP}$ and $\text{Rec} = \frac{TP}{TP + FN}$. Here, $TP$, $FP$, and $FN$ denote true positives, false positives, and false negatives, respectively.  
(2) \emph{Tracking Accuracy}: 
We report \textbf{AMOTA} and \textbf{AMOTP} following
\cite{weng2020ab3dmot, bernardin2008evaluating}.
AMOTA reflects trajectory-level tracking quality, capturing whether the target can be \emph{continuously} recovered over time while penalizing missed targets, false detections, and identity switches.
AMOTP reflects the localization precision of the matched tracking boxes, measuring how \emph{accurately} the recovered boxes align with the target positions.
(3) \emph{Efficiency}: We measure retrieval latency as the per-query operator execution time. For tracking, we report latency per 100 processed scenes
using wall-clock time.

\noindent
\textbf{Evaluation Platform}. The experiments are conducted on a 10-core server with  Intel(R) Core(TM) i9-9820X CPU @ 3.30GHz, 64GB RAM, and 2 NVIDIA GeForce RTX 2080 Ti GPUs (11GB each).

\subsection{Experimental Results}
\label{sec:emma_exp_result}

\noindent
\textbf{RQ1. Retrieval Evaluation.} We evaluate the performance of retrieval operator.

\noindent
\emph{(1) Retrieval performance evaluation.} The results of retrieval performance across different datasets are reported in Table~\ref{tab:detection_evaluation}. 
\textit{Call Rt.} denotes the average number of model invocations per scene, while \textit{Joint Rt.} denotes the fraction of model-processing steps that use joint analysis.
To summarize the accuracy-efficiency trade-off, we define \texttt{F1-Eff.} as $F1\cdot E_{\mathrm{norm}}$, where $E_{\mathrm{norm}}=\frac{L_{\min}}{L}$, $L$ denotes the method latency, and $L_{\min}$ is the minimum latency across all methods within the same dataset block.

\begin{table}[ht]
    \centering
    \caption{Evaluation on retrieval workload.}
    \label{tab:detection_evaluation}
    \vspace*{-2.5mm}
    \scriptsize
    \setlength{\tabcolsep}{2.6pt}
    \renewcommand{\arraystretch}{0.9}
    \resizebox{\columnwidth}{!}{
    \begin{tabular}{c |c| l| c| c | c| c| c}
    \hline
     \textbf{Data} & $|\mathcal{Q}|$ & \textbf{Method} & \textbf{Call Rt.}
     & \textbf{Joint Rt.} & \textbf{F1} & \textbf{Avg Lat. (min)} & \textbf{F1-Eff.} \\
     \hline
     \multirow{6}{*}{OPV2V} & \multirow{6}{*}{110} & PointPillar & 1.000 & 0.000 & 0.836 & \textbf{1.11} & 0.836 \\
      &  & VoxelNet-Near & 1.131 & 1.000 & 0.943 & 10.54 & 0.099  \\
      &  & VoxelNet-All & 3.054 & 1.000 & 0.955 & 11.66 & 0.091  \\
      &  & MAST & 0.269 & 1.000 & 0.929 & 2.50 & 0.412  \\
      &  & EMMA-H & 0.519 & 0.218 & 0.869 & 1.58 & 0.611  \\
      &  & EMMA & 0.250 & 0.562 & 0.931 & 1.44 & 0.716  \\
     \hline
     \multirow{6}{*}{DOMV} & \multirow{6}{*}{720} & PointPillar & 1.000 & 0.000 & 0.461 & 20.46 & 0.087  \\
      &  & VoxelNet-Near & 1.003 & 1.000 & 0.816 & 172.06 & 0.018  \\
      &  & VoxelNet-All & 8.996 & 1.000 & 0.885 & 632.39 & 0.005  \\
      &  & MAST & 0.102 & 1.000 & 0.778 & 17.52 & 0.172  \\
      &  & EMMA-H & 0.265 & 0.035 & 0.828 & 13.13 & 0.244  \\
      &  & EMMA & 0.096 & 0.108 & 0.878 & \textbf{3.87} & \textbf{0.878}  \\
     \hline
     \multirow{5}{*}{DOMV-Long} & \multirow{5}{*}{20} & PointPillar & 1.000 & 0.000 & 0.173 & 408.8 & 0.091  \\
      &  & VoxelNet-Near & 1.001 & 1.000 & 0.454 & 3431.2 & 0.029  \\
      &  & MAST & {0.109} & 1.000 & 0.495 & 372.1 & 0.288  \\
      &  & EMMA-H & 0.303 & 0.142 & 0.471 & 1811.0 & 0.056  \\
      &  & EMMA & 0.164 & 0.358 & 0.512 & \textbf{216.4} & \textbf{0.51}  \\
     \hline
    \end{tabular}
    }
    \renewcommand{\arraystretch}{1.0}
\end{table}

As shown in Table~\ref{tab:detection_evaluation}, on DOMV, EMMA provides the best accuracy-efficiency trade-off.
Compared with the adapted MAST, a competitive baseline in terms of F1-Eff., EMMA improves F1 by $12.9\%$ ($0.878$ vs. $0.778$) and reduces latency by $4.5\times$ ($17.52$ vs. $3.87$ minutes).
This is because EMMA reduces the model-call rate ($0.096$ vs. $0.102$) and joint-analysis rate ($0.108$ vs. $1.000$).
EMMA-H also achieves a higher F1-Eff. score compared to MAST.
EMMA further improves over EMMA-H by $0.050$ F1 while being $3.4\times$ faster, showing the benefit of learned policies over heuristic rules.
Compared with VoxelNet-All, the most accurate but also the most computationally expensive baseline, EMMA is $163.4\times$ faster than VoxelNet-All while still preserving a high F1 of $0.878$.

On OPV2V, PointPillar achieves the highest F1-Eff.\ ($0.836$). This is because that PointPillar is pretrained on OPV2V, and ego-only inference already achieves an F1 of $0.836$ at the lowest average latency ($1.11$ minutes per query).
Compared with DOMV, EMMA achieves a smaller accuracy-efficiency gain on OPV2V.
In the OPV2V retrieval workload, each ego vehicle has at most two neighboring sensor-equipped vehicles available for joint analysis, and their views overlap relatively consistently. This leaves fewer opportunities for EMMA to improve retrieval by selecting a complementary viewpoint.
EMMA's policies are trained on DOMV and applied to OPV2V without fine-tuning.
Even in this setting, EMMA raises F1 from $0.836$ to $0.931$ over PointPillar, an $11.4\%$ relative improvement, while increasing average latency from $1.11$ to $1.44$ minutes, or about $20$ seconds per query.
EMMA achieves an F1 score comparable to the adapted MAST
($0.931$ vs.\ $0.929$).
EMMA reduces average latency from $2.50$ to $1.44$ minutes and achieves the highest F1-Eff.\ among the methods that use multi-vehicle observations ($0.716$).
EMMA-H also achieves a higher F1-Eff.\ score than MAST
($0.611$ vs.\ $0.412$) by reducing latency.

DOMV-Long further evaluates scalability under long-sequence retrieval.
VoxelNet-All is omitted on DOMV-Long because its average per-query latency exceeds one day and times out in our evaluation.
EMMA achieves an F1 of $0.512$ with $216.4$ minutes of latency, outperforming VoxelNet-Near by $12.8\%$ F1 while being $15.9\times$ faster.
It outperforms MAST by $3.4\%$ F1 with a $1.7\times$ latency reduction, and outperforms EMMA-H by $8.7\%$ F1 while being $8.4\times$ faster.
The speedup over EMMA-H increases from $3.4\times$ on DOMV to $8.4\times$ on DOMV-Long, highlighting the benefit of learned policies for long-sequence retrieval.

\smallskip
\noindent
\emph{(2) Evaluation by varying cardinality.}
The results for the retrieval workload with varying query cardinalities are shown in Fig.~\ref{fig:detection_vary_cardinality}.

\begin{figure}[ht]
    \centering
    \includegraphics[width=0.84\linewidth]{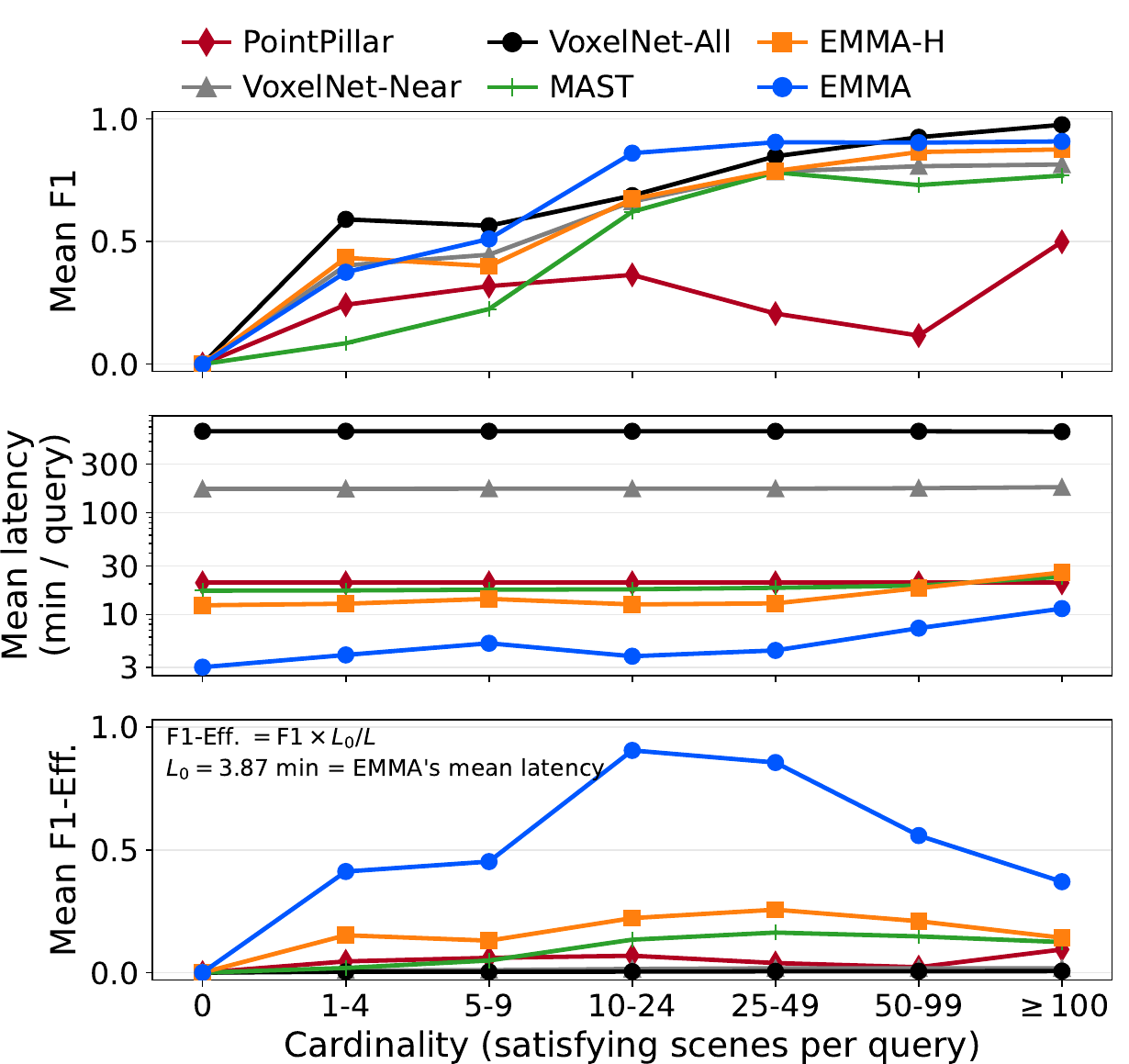}
    \vspace*{-2.5mm}
    \caption{Retrieval performance under varying cardinalities.}
    \vspace*{-2.5mm}
    \label{fig:detection_vary_cardinality}
\end{figure}

EMMA's
F1 is higher when more scenes satisfy the query than when the query is highly selective. This is because fine search helps EMMA recover consecutive value scenes once the predicate starts to hold. 
In contrast, 
MAST struggles in the highly selective region, with its F1 staying 
at a low level
for queries with cardinality 
$0$ to $9$.
The latency trend also reflects EMMA's adaptive behavior. For selective queries, most scenes are irrelevant, so EMMA skips invalid regions and finishes faster. For larger cardinality, EMMA performs finer analysis to recover more matched scenes.

\smallskip
\noindent
\emph{(3) Retrieval with multi-modal predicates.}
{We further evaluate retrieval queries with multi-modal predicates on DOMV, where the query contains both 3D geometric constraints and visual color constraints.}
{Since these predicates require point-cloud detection followed by image-based color verification, we report the results separately in Table~\ref{tab:retrieval_multimodal}, where Add. denotes additional.}

\begin{table}[ht]
    \centering
    \vspace*{-2mm}
    \caption{Retrieval with multi-modal predicates.}
    \label{tab:retrieval_multimodal}
    \vspace*{-2.5mm}
    \scriptsize
    \setlength{\tabcolsep}{3.0pt}
    \renewcommand{\arraystretch}{0.9}
    \resizebox{.95\columnwidth}{!}{
    \begin{tabular}{c|c|c|c|c|c}
    \hline
     \textbf{Data} & $|\mathcal{Q}|$ & \textbf{Predicate} & \textbf{Method} & \textbf{F1} & \textbf{Add. Avg Lat. (s)} \\
     \hline
     \multirow{6}{*}{DOMV} & \multirow{6}{*}{100} & \multirow{6}{*}{Color} & PointPillar & 0.065 & +2.4 \\
      &  &  & VoxelNet-Near & 0.446 & +12.6 \\
      &  &  & VoxelNet-All & 0.475 & +16.6 \\
      &  &  & MAST & 0.392 & +11.6 \\
      &  &  & EMMA-H & 0.411 & +12.0 \\
      &  &  & EMMA & \textbf{0.419} & +14.3 \\
     \hline
    \end{tabular}
    }
    \renewcommand{\arraystretch}{1.0}
    \vspace*{-1mm}
\end{table}

{We use CLIP ViT-L~\cite{radford2021learning} as the image model for color verification.}
{Overall, the F1 scores in Table~\ref{tab:retrieval_multimodal} roughly follow and are bounded by the geometric retrieval performance, because visual verification is applied after the geometric retrieval stage.}
{In our current setting, CLIP ViT-L achieves about $50\%$ color recognition accuracy, and our empirical study shows that occlusion can substantially reduce its correctness.}
{The additional latency is also consistent with the number of CLIP invocations, i.e., the number of scenes sent to the image model for color verification.}
\bluecolor{These results show that the accuracy of visual predicate evaluation
depends on the underlying visual model.}

\smallskip
\noindent
\textbf{RQ2. Tracking Evaluation.}
The experimental results of tracking performance are reported in Table~\ref{tab:tracking_evaluation}.

\begin{table}[ht]
    \centering
    \caption{Evaluation on tracking operator.}
    \label{tab:tracking_evaluation}
    \vspace*{-2.5mm}
    \scriptsize
    \setlength{\tabcolsep}{3.5pt}
    \renewcommand{\arraystretch}{0.9}
    \resizebox{.95\columnwidth}{!}{
    \begin{tabular}{c|c|c|c|c|c}
    \hline
     \textbf{Dataset} & $|\mathcal{Q}|$ & \textbf{Method} & \textbf{AMOTA} & \textbf{AMOTP} & \textbf{Avg Lat. (s/100 frames)}\\
     \hline
     \multirow{5}{*}{OPV2V} & \multirow{5}{*}{100} & PointPillar & 0.884 & 0.177 & 13.32 \\
      &  & VoxelNet-Near & 0.931 & 0.132 & 69.23  \\
      &  & VoxelNet-All & 0.984 & 0.136 & 76.51  \\
      &  & EMMA-H & 0.939 & 0.196 & 24.52  \\
      &  & EMMA & 0.962 & 0.198 & \bluecolor{59.88}  \\
     \hline
     \multirow{5}{*}{DOMV} & \multirow{5}{*}{258} & PointPillar & 0.331 & 0.259 & 9.93 \\
      &  & VoxelNet-Near & 0.481 & 0.246 & 90.63 \\
      &  & VoxelNet-All & 0.880 & 0.237 & 140.37  \\
      &  & EMMA-H & 0.373 & 0.278 & 31.76  \\
      &  & EMMA & 0.775 & 0.286 & \bluecolor{62.11} \\
     \hline
     \multirow{4}{*}{DOMV-Long} & \multirow{4}{*}{20} & PointPillar & 0.504 & 0.227 & 9.95 \\
      &  & VoxelNet-Near & 0.724 & 0.152 & 90.82 \\
      &  & EMMA-H & 0.653 & 0.211 & 26.61  \\
      &  & EMMA & 0.910 & 0.293 & \bluecolor{58.28} \\
     \hline
    \end{tabular}
    }
    \renewcommand{\arraystretch}{1.0}
    \vspace*{-1mm}
\end{table}

For tracking, AMOTA is the primary accuracy metric because it reflects how completely the target trajectory is recovered over its ground-truth duration, while AMOTP measures whether the recovered bounding boxes are accurate in terms of location.
On DOMV, PointPillar performs poorly with an AMOTA of $0.331$, because a missed detection at one frame can break the target association and truncate the entire trajectory.
VoxelNet-All achieves the best tracking accuracy with an AMOTA of $0.880$ and AMOTP of $0.237$, but it requires $140.37$ seconds per $100$ frames.
EMMA reduces the latency to $62.11$ seconds, \bluecolor{providing a $2.26\times$ speedup over VoxelNet-All}, while still preserving a high AMOTA of $0.775$.
EMMA also outperforms VoxelNet-Near in AMOTA ($0.775$ vs.\ $0.481$), showing that the policy can effectively select useful collaborative perception pairs rather than relying on a fixed nearby source.

On DOMV-Long, VoxelNet-All is also omitted due to the evaluation budget.
Among the reported methods, VoxelNet-Near achieves the best AMOTP ($0.152$), indicating accurate localization for the matched boxes, while EMMA achieves the best AMOTA ($0.910$), indicating stronger trajectory-level recovery, and \bluecolor{runs $1.56\times$ faster than VoxelNet-Near}.
This result shows that the most expensive joint perception choice is not always the best for tracking continuity: in rare cases, an independent action can keep the target associated while a joint action fails.
{By selecting between independent and joint actions at each tracking step, EMMA can therefore obtain higher AMOTA than fixed collaborative execution while still maintaining moderate latency.}

On OPV2V, \bluecolor{EMMA provides a smaller latency reduction over VoxelNet-All ($1.28\times$)}.
Similar to retrieval, this is because OPV2V sequences are short, and the relative geometry between vehicles is more stable.
As a result, VoxelNet-Near and VoxelNet-All already achieve similar AMOTA values ($0.931$ and $0.984$), leaving less room for policy-guided source selection.
These results show that EMMA's advantage grows with scene stream length and changes in the overlap among vehicle observations.

\noindent
\textbf{RQ3. Composed Workload Evaluation.}
We evaluate composed retrieval-and-tracking workloads where retrieved target objects are tracked on the
DOMV dataset.
The goal is to measure composed-operator performance under different levels of shared information: w/o shared cache, w/ Intra-Query Cache, where cache is shared across operators within one query, and w/ Inter-Query Cache, where cache is further shared across queries.
The results are shown in Table~\ref{tab:tracking_evaluation_composed}.

\begin{table}[ht]
    \centering
    \caption{Evaluation on composed tracking workload.}
    \label{tab:tracking_evaluation_composed}
    \vspace*{-2.5mm}
    \scriptsize
    \setlength{\tabcolsep}{3.5pt}
    \renewcommand{\arraystretch}{0.92}
    \resizebox{\columnwidth}{!}{
    \begin{tabular}{l|c|c|c}
    \hline
     \textbf{Metric} &
     \textbf{w/o Shared Cache} &
     \textbf{w/ Intra-Query Cache} &
     \textbf{w/ Inter-Query Cache} \\
     \hline \hline
     AMOTA & 0.689 & 0.689 & 0.689 \\
     AMOTP & 0.344 & 0.344 & 0.344 \\
     Ind Calls & 905.5 & 898.4 & 421.4 \\
     Joint Calls & 1006.4 & 999.7 & 690.6 \\
     Retrieval Lat. & 266.4 & 266.4 & \textbf{222.9} \\
     Tracking Lat. & \bluecolor{1510.0} & \bluecolor{1503.2} & \bluecolor{\textbf{1230.6}} \\
     Avg Lat. & \bluecolor{1776.4} & \bluecolor{1769.6} & \bluecolor{\textbf{1453.5}} \\
     \hline
    \end{tabular}
    }
    \renewcommand{\arraystretch}{1.0}
    \vspace*{-1mm}
\end{table}

Across the three variants, the composed query results remain the same; in particular, the tracking AMOTA and AMOTP are unchanged. Intra-query sharing only slightly reduces the downstream tracking latency, because the cache produced by a single retrieval query provides limited reusable evidence for the following tracking operator. In contrast, inter-query sharing accumulates reusable world-store information across multiple queries on the same data, reducing retrieval latency by 16.3\% (from 266.4 to 222.9 seconds) and tracking latency by 18.5\% (from 1510.0 to 1230.6 seconds).

\noindent
\textbf{RQ4. Shared Cache Evaluation.}
We evaluate sharing the world store across consecutive retrieval queries over the
DOMV dataset.
Previously detected objects and 
extracted information 
can be reused by later queries, reducing redundant perception calls.

\begin{figure}[ht]
    \centering    
    \vspace{-2mm}
    \includegraphics[width=0.7\linewidth]{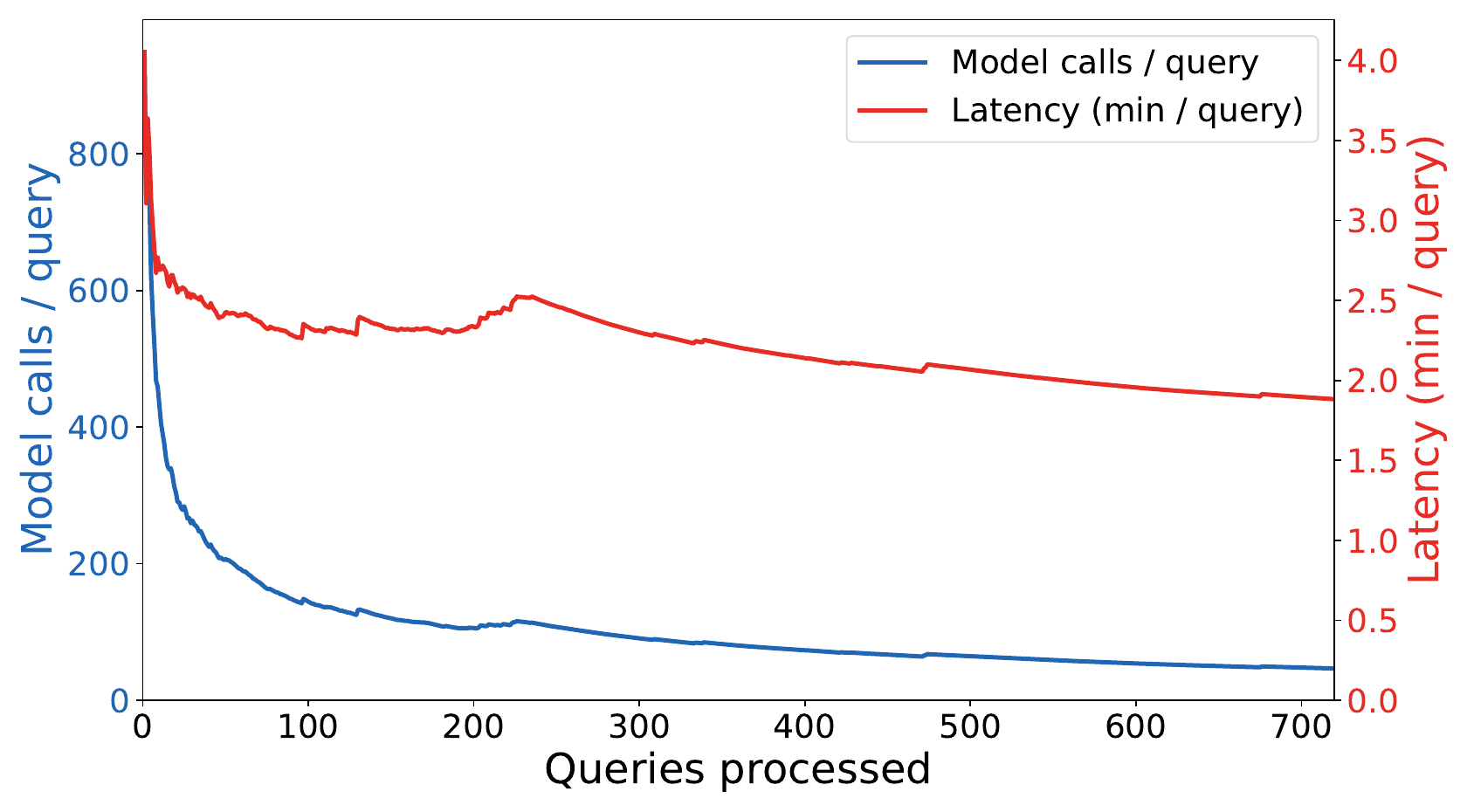}
    \vspace*{-3mm}
    \caption{Shared world-store reuse on 
    retrieval queries.}
    \vspace{-3mm}
    \label{fig:shared_world_store}
\end{figure}

As shown in Fig.~\ref{fig:shared_world_store}, both latency and the number of model predictions are relatively high at the beginning, when the world store contains little reusable evidence.
During the first 200 queries, the latency and model calls drop rapidly but still fluctuate, because some queries touch scenes that have not yet been processed by the perception models.
After roughly 200 queries, both curves decrease more steadily: most frequently needed scene evidence has already been cached, and later queries only need to invoke models on a smaller set of unseen or uncertain scenes.
Compared with $3.87$ min/query without reuse, world-store reuse reduces latency to $2.333$ and $2.068$ min/query after $100$ and $500$ queries ($1.66\times$ and $1.87\times$ speedups), and to $1.883$ min/query after all $720$ queries ($2.06\times$).

\noindent
\textbf{RQ5. Runtime Evaluation.}
The runtime results of each component in our system are reported in
Fig.~\ref{fig:runtime}, including retrieval and tracking queries.

\begin{figure}[ht]
    \centering
    \vspace{-3mm}
    \makebox[\linewidth][c]{%
        \includegraphics[width=0.95\linewidth]{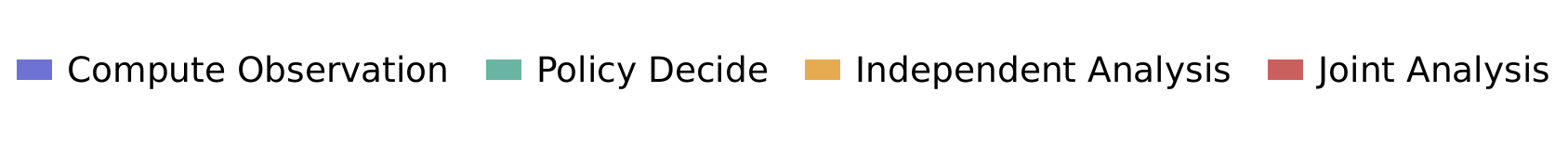}%
    }\par
    \vspace{-2.5mm}
    \begin{subfigure}[t]{0.49\linewidth}
        \centering
        \includegraphics[width=\linewidth, trim=0 15pt 0 0, clip]{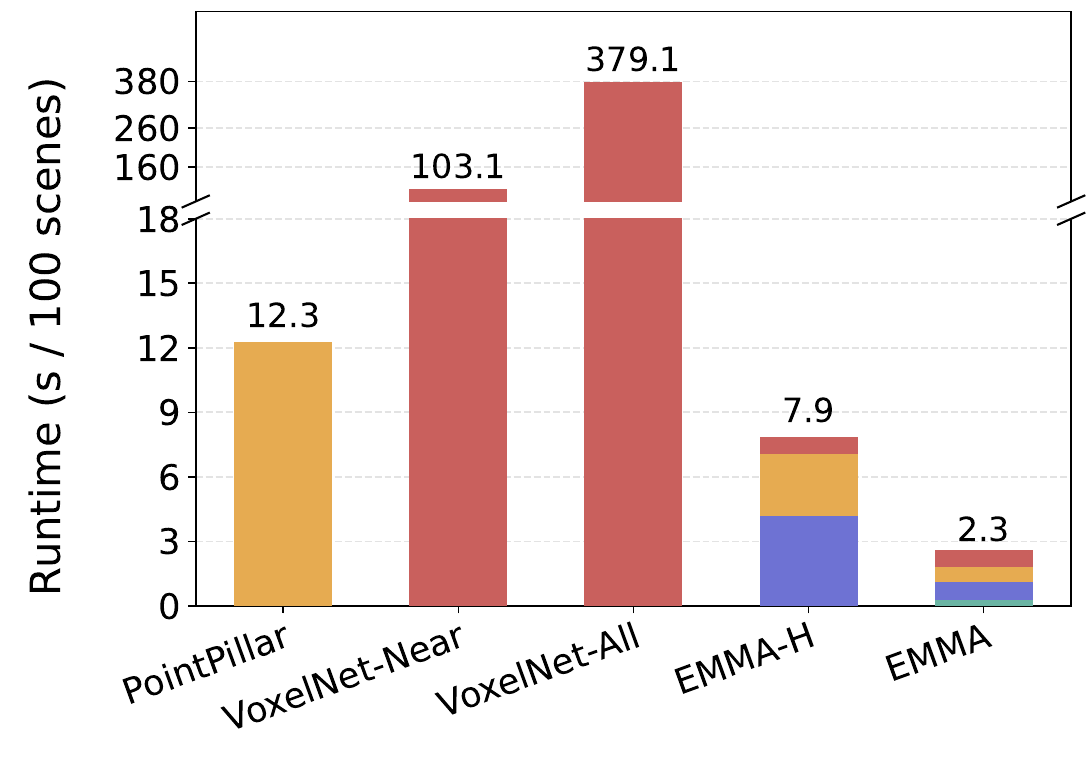}
        \caption{Retrieval Query}
        \label{fig:retrieval_runtime}
    \end{subfigure}
    \hfill
    \begin{subfigure}[t]{0.49\linewidth}
        \centering
        \includegraphics[width=\linewidth, trim=0 15pt 0 0, clip]{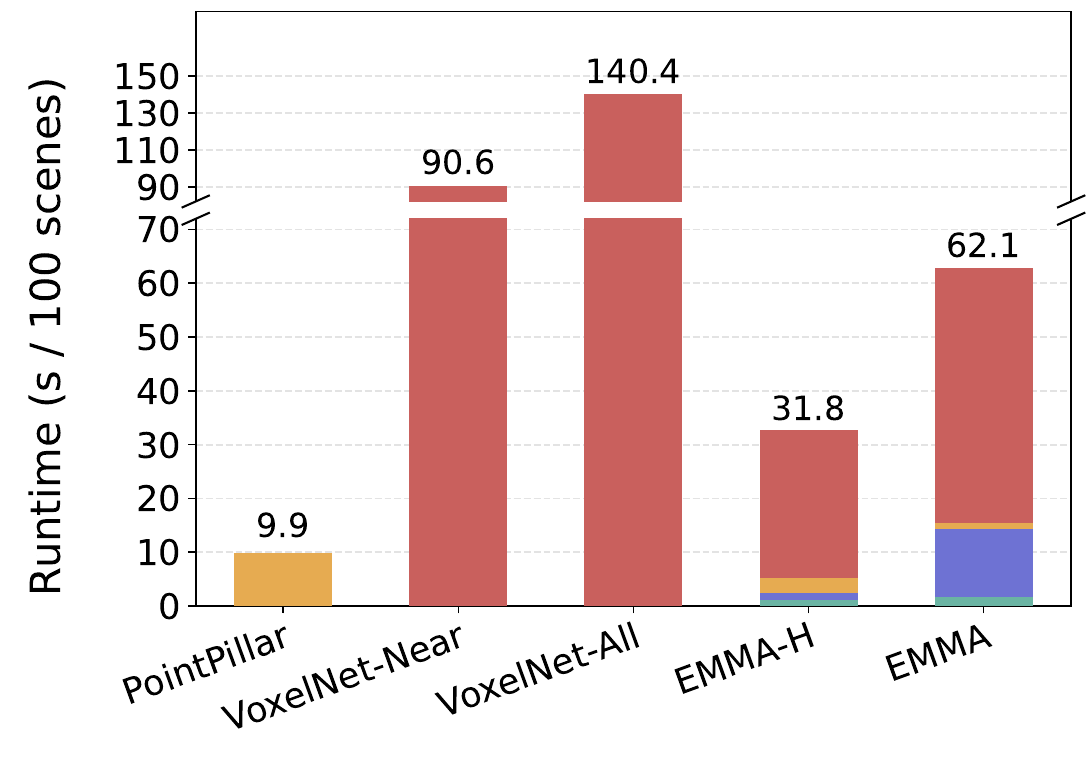}
        \caption{Tracking Query}
        \label{fig:tracking_runtime}
    \end{subfigure}
    \vspace*{-2mm}
    \caption{Runtime breakdown of each component for retrieval and tracking queries.}
    \label{fig:runtime}
\end{figure}

As shown in Fig.~\ref{fig:runtime}, PointPillar and VoxelNet spend almost all runtime on deep model processing in both retrieval and tracking, because they do not include policy-guided observation construction or source selection.
For retrieval, EMMA-H uses independent analysis more often and relies less on joint analysis, while EMMA selects joint analysis more frequently on processed scenes but keeps total model usage low through effective temporal skipping.
\bluecolor{For tracking, EMMA incurs a higher observation-construction cost
because it applies a Kalman filter 
to predict the target's
current bounding box.}

\noindent
\textbf{RQ6. Retrieval Skip-Range Evaluation.}
\bluecolor{Figure~\ref{fig:retrieval_skip_pareto} reports the results under
different scene-skipping ratios, where a higher ratio indicates more
aggressive skipping.}
We use query throughput, computed as $1/L$ where $L$ is query latency in minutes, on the x-axis to compare accuracy-efficiency trade-offs.

\begin{figure}[ht]
    \centering
    \includegraphics[width=0.7\linewidth]{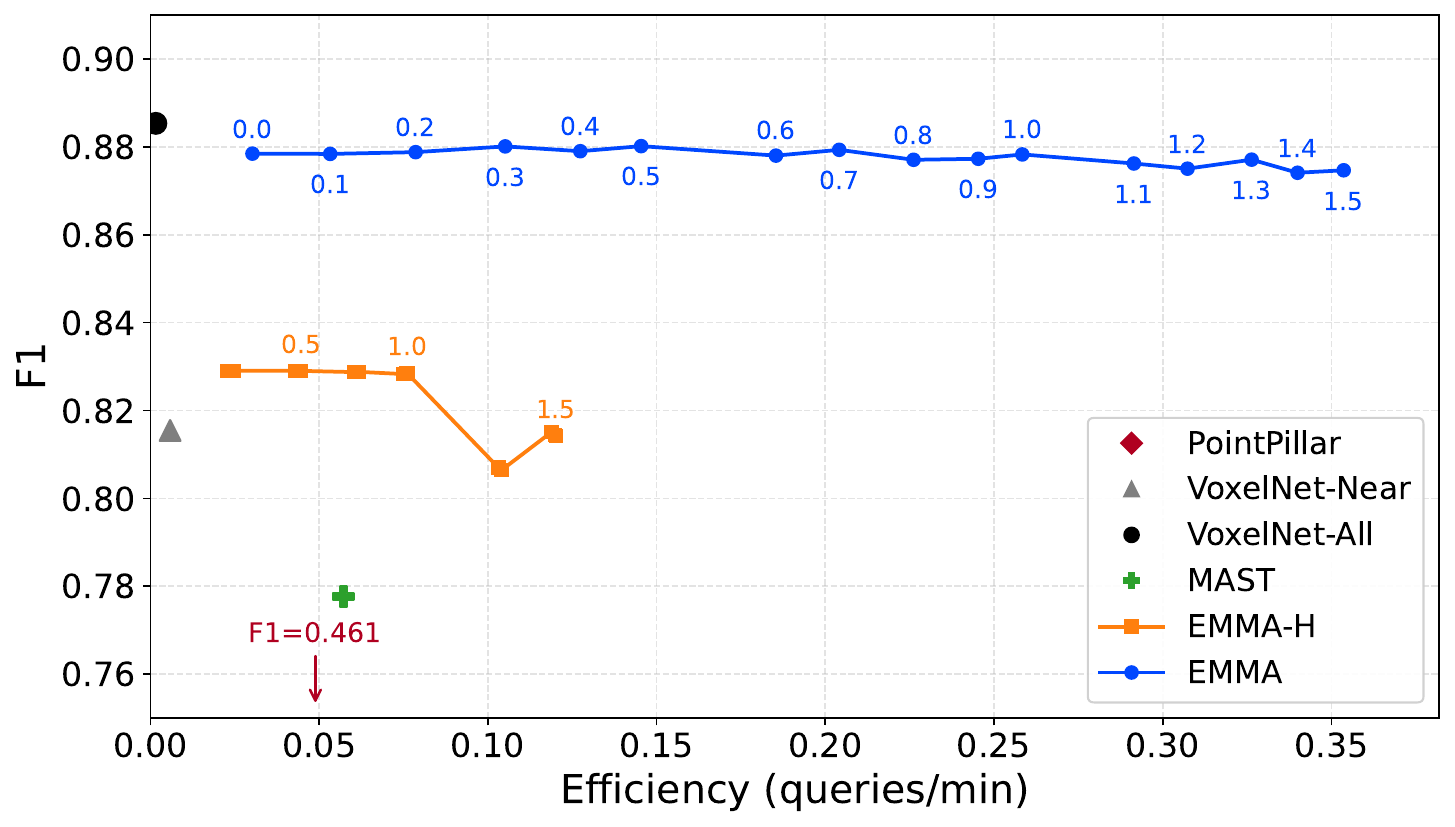}
    \vspace*{-2mm}
    \caption{Accuracy-efficiency trade-off by varying scene skipping ratio.}
    \vspace*{-3mm}
    \label{fig:retrieval_skip_pareto}
\end{figure}

As shown in Fig.~\ref{fig:retrieval_skip_pareto}, EMMA remains robust under different skipping ratios. Although its query throughput varies as the temporal policy changes how aggressively frames are skipped, its F1 score stays largely stable, indicating that the learned policy can avoid skipping most value frames. In contrast, EMMA-H loses F1 more quickly when its efficiency improves.
Both EMMA and EMMA-H provide better accuracy-efficiency trade-offs than MAST.
The other two  baselines  illustrate the two extremes: PointPillar is fast but has very low retrieval accuracy, while VoxelNet-All achieves high F1 only with prohibitive latency. These results further validate the benefit of learning an adaptive temporal policy for retrieval execution.

\noindent
\textbf{RQ7. Training Curve.} The training curves of retrieval operator policies are shown in Fig.~\ref{fig:retrievaltemporal} and~\ref{fig:retrievaldecide}. 
The tracking policy training curves are provided in Appendix~\ref{app:tracking_training_curves} (Fig.~\ref{fig:tracking_training_curves}).
\begin{figure}[ht]
    \centering
    \begin{subfigure}[t]{0.48\linewidth}
        \centering
        \includegraphics[width=\linewidth, trim=0 10pt 0 0, clip]{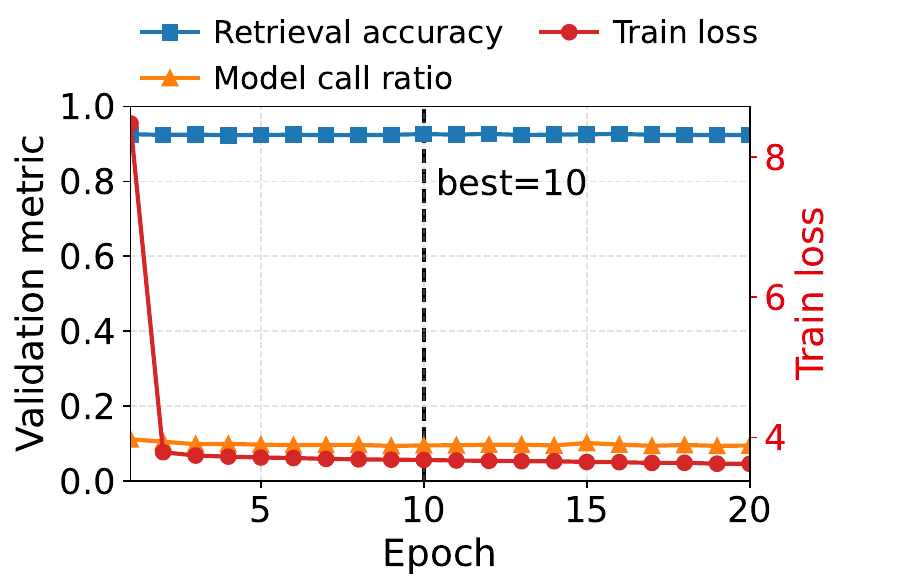}
        \caption{PolicyTemporal}
        \label{fig:retrievaltemporal}
    \end{subfigure}
    \hfill
    \begin{subfigure}[t]{0.48\linewidth}
        \centering
        \includegraphics[width=\linewidth, trim=0 10pt 0 0, clip]{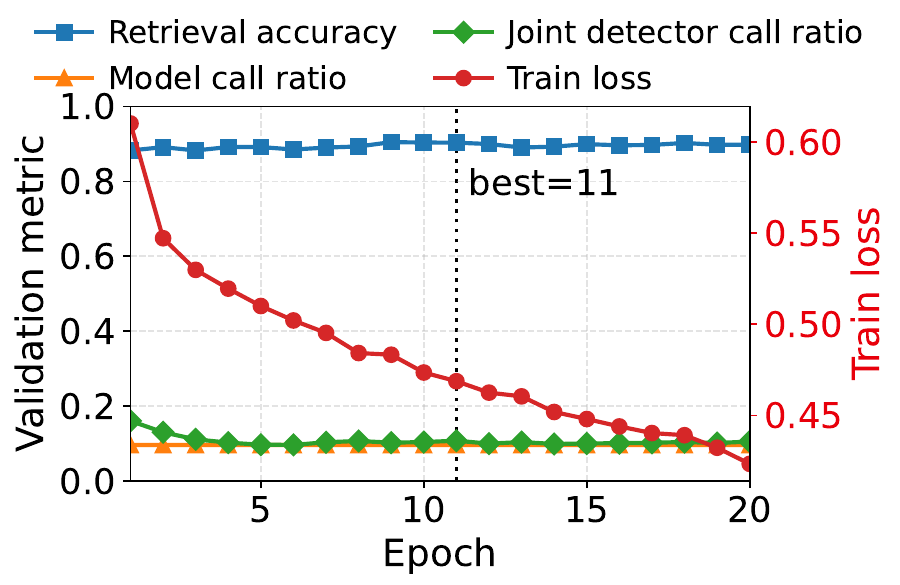}
        \caption{PolicyDecide}
        \label{fig:retrievaldecide}
    \end{subfigure}
    \vspace*{-2.5mm}
    \caption{Retrieval policy training curves.}
    \label{fig:retrieval_training_curves}
    \vspace*{-2mm}
\end{figure}

As shown in Fig.~\ref{fig:retrieval_training_curves}, the retrieval policies reach their best observed validation performance at similar epochs: \textsc{PolicyTemporal} at epoch 10 and \textsc{PolicyDecide} at epoch 11.
For tracking, \textsc{PolicyDecide} reaches its best observed validation performance later, at epoch 60, whereas \textsc{PolicyTemporal} reaches its best at epoch 9, close to its retrieval counterpart.
Retrieval supervision is derived directly from ground-truth scenes satisfying the query predicates, whereas tracking supervision is generated from trajectory rollouts whose association outcomes depend on earlier decisions in each rollout.

\section{Related Work}
\label{sec:emma_related_work}

\smallskip
\noindent
\textbf{Multi-Modal Data Analytics.}
Vehicles and robots record high-volume multi-modal data, including point clouds,
video, and trajectories. Prior work has studied data management and analytical
query processing for these modalities, including unified representations for
self-driving logs~\cite{segal2021universal}.
{Point-cloud management has been studied for map-level spatial queries and
object-level similarity search~\cite{yan2024proximity, liu2025efficient}.}
For sequential vehicle LiDAR data, MAST~\cite{li2025mast, li2025mast-demo}
combines spatial and semantic filters over point-cloud sequences and uses
reinforcement-learning-based importance sampling to improve query efficiency.
{Beyond LiDAR streams, prior work has studied accurate video understanding and retrieval over camera streams~\cite{huang2022multi}, as well as database-oriented video or image data storage, querying, and visual analysis interfaces~\cite{kittivorawong2024spatialyze, ge2021demonstration, DBLP:journals/pacmmod/WuCPCAR24, sirin2025data, bang2023seiden}.}
{Prior work has also studied trajectory analysis and management~\cite{yuan2026goodtp, xie2022ganos}.}

Existing systems primarily optimize individual modalities or recordings. They
do not support composed retrieval and tracking functionalities over  multi-modal
observations 
\bluecolor{whose spatial overlap changes over time as
the recording vehicles or robots move.}

\smallskip
\noindent
\textbf{Cooperative Vehicle Perception.}
Prior work on cooperative vehicle perception combines observations, such as
point-cloud frames, from different vehicles to improve predefined tasks such
as object detection~\cite{shi2020pv} and trajectory
prediction~\cite{ruan2024learning,DBLP:conf/icra/XuXXHLM22}.
Fusion methods~\cite{huang2022multi, DBLP:conf/icra/XuXXHLM22} are typically categorized into three types: early fusion, 
which merges raw point clouds from multiple vehicles to produce a combined point cloud with extended range and/or higher density;
intermediate fusion, which combines feature embeddings extracted from individual point clouds for joint prediction; and late fusion, which aggregates high-level information such as bounding boxes extracted from multiple vehicles for downstream analysis.

These methods primarily optimize predefined perception tasks using a fixed
fusion pipeline and mainly focus on accuracy improvement. EMMA instead treats whether and how recorded observations
should be jointly analyzed as a query-execution decision that can change
across queries and scenes.

\smallskip
\noindent
\textbf{Multi-Perspective Video Analytics.}
{Recent systems have studied data management and query processing for multiple geospatial video streams. Spatialyze~\cite{kittivorawong2024spatialyze} uses spatial-aware optimizations to prune irrelevant frames, while Apperception~\cite{ge2021demonstration} and VisualWorldDB~\cite{DBLP:conf/cidr/HaynesDMBCC20} organize multi-perspective videos into a common world abstraction for object and trajectory queries.}
\bluecolor{Unlike these video-centered systems, EMMA coordinates processing
across video, LiDAR point clouds, and GPS trajectories within a single
query plan. It stages modality-specific predicate evaluation, adapts
independent or joint analysis as observation overlap changes across
scenes, and reuses extracted object states across composed retrieval
and tracking operators.}

\section{Conclusion}
\label{sec:emma_conclusion}

In this paper, we present EMMA, an operator-oriented framework
for queries over historical multi-modal data recorded by vehicles
and robots. EMMA compiles queries into composable 
retrieval
and tracking
operators and invokes deep models on demand. It stages
modality-specific predicate evaluation and combines overlapping
observations when 
data from one vehicle/robot
is insufficient to achieve high query accuracy.
Offline-trained plug-in policies select independent or joint analysis
and safe scene-skip lengths from lightweight observations. A
co-location index prunes joint-analysis candidates, and a shared world
store reuses extracted object states. Experiments show that EMMA
substantially reduces query cost while preserving accuracy.
We plan to release DOMV and DOMV-Long, along with the retrieval and tracking workloads developed for this study.

\clearpage
\appendix

\section{Example Python Workflow}
\label{app:python_workflow}

Listing~\ref{lst:example-python} illustrates the Python workflow corresponding to the declarative query in Listing~\ref{lst:application-query-new}.
The workflow constructs an environment from the selected agents, restores aligned data, executes scene retrieval, and then tracks the retrieved target objects.

\begin{lstlisting}[
style=ummaPython,
caption={EMMA-Plan workflow.},
label={lst:example-python},
numbers=left
]
env = Environment()
env.add_agents(['V_A', 'V_B', 'V_C'])
env.restore(interval=('10:00', '11:00'))  # synchronize and align data

scenes = DetectScenes(
    env,
    interval=('10:00', '11:00'),
    predicates=[
        ('car_1',
         Color('red'),
         Near(Point3D(x1, y1, z1), r1),
         Speed(v1_min, v1_max)),
        ('car_2',
         Color('green'),
         Near(Point3D(x2, y2, z2), r2),
         Speed(v2_min, v2_max)),
    ]
)

seeds = scenes.objects(label='car_1')
trajectories = Track(env, targets=seeds)
\end{lstlisting}

Listing~\ref{lst:example-python} illustrates the execution workflow of our system, represented in Python.
First, the system constructs an environment object {\tt env} based on the three agents {\tt [V\_A, V\_B, V\_C]} (lines 1--2).
These agents are managed within {\tt env}, and their data are preprocessed (e.g., temporally and spatially aligned) before query execution (line 3).
The workflow then performs a detection operation on the constructed environment (lines 5--18).
Rather than exhaustively processing all video and point cloud streams for detection, the system employs several optimizations, including multi-stage pruning mechanisms, to improve query efficiency.
In particular, during the execution of {\tt DetectScenes}, the intermediate outputs from deep models are cached in {\tt env} and reused by subsequent operations such as {\tt Track}.

\section{Sub-Operator Pseudocode}

\begin{algorithm}[htbp]
\caption{Sub-Operators of EMMA.}
\label{alg:policy_library}
\small
\DontPrintSemicolon
\SetKwProg{Fn}{Function}{:}{}

\Fn{\textsc{Observe}$(S,t,\mathcal{W},\texttt{task})$}{
  \tcp{Construct task-aware lightweight observation}
  $\texttt{hint} \leftarrow \mathcal{W}.lookup(\textsc{ID}(S),t)$\;
  $\texttt{obs} \leftarrow$ task-aware scene context at time $t$ of $S$ with $\texttt{hint}$\;
  \lIf{{$\texttt{task}=\texttt{tracking}$}}{{$\texttt{obs} \leftarrow \texttt{obs}\cup\{\textsc{KFPredict}(\texttt{hint})\}$};}
  \Return $\texttt{obs}$\;
}

\Fn{\textsc{PolicyDecide}$(\pi,\texttt{task},\texttt{pred},\texttt{obs})$}{
  \tcp{{Policy maps the state to an analysis action}}
  $\mathbf{a} \leftarrow \pi(\texttt{task},\texttt{pred},\texttt{obs})$\;
  \Return $\mathbf{a}$\;
}

\Fn{\textsc{Execute}$(S, t, \mathbf{a},\texttt{obs})$}{
  \tcp{Execute the pipeline specified by action $\mathbf{a}$}
  $\texttt{feat} \leftarrow$ features extracted from $\texttt{obs}$ and $S(t)$ according to $\mathbf{a}$\;
  \Return $\texttt{feat}$\;
}

\Fn{{\textsc{Track}$(b,\texttt{obs},\texttt{feat})$}}{
  \tcp{{Update state $b$ using extracted tracking features}}
  {$b' \leftarrow$ refined target state estimated from $(b,\texttt{obs},\texttt{feat})$}\;
  \Return $b'$\;
}

\Fn{\textsc{Eval}$(\varphi,S,t,\texttt{feat})$}{
  $m \leftarrow$ \textsc{BoxHit}$(\varphi,\texttt{feat})$; \tcp{Box-level hit}
  \tcp{Optional speed/color/shape filters}
  \If{$m$ \textbf{and} \textsc{HasVisualPred}$(\varphi)$}{
  $m \leftarrow
  \textsc{VisualVerify}(\varphi,S(t),\texttt{feat})$\;
    }
  \Return $m$\;
}

\Fn{\textsc{SyncWorld}$(\mathcal{W},S,t, \mathbf{a} ,\texttt{feat})$}{
  $\mathcal{W}[\langle \textsc{ID}(S),t,\mathbf{a}]
  \leftarrow \texttt{feat}$\;
  \Return $\mathcal{W}$\;
}

\Fn{\textsc{PolicyTemporal}$(\pi_{\text{temp}},\texttt{task},\chi_t)$}{
  $\mathbf{u} \leftarrow \pi_{\text{temp}}(\texttt{task},\chi_t)$; \tcp{{Select a temporal action}} \
  $(\texttt{out}_{\Delta},\texttt{aux}_{\Delta},t') \leftarrow \textsc{TemporalExe}(\texttt{task},\mathbf{u},\chi_t)$\;
  \Return~$(\texttt{out}_{\Delta},\texttt{aux}_{\Delta},t')$\;
}

\end{algorithm}

\section{Policy Observation Details}
\label{app:policy_observation_details}
This section details the lightweight observations used by the policy modules in Section~\ref{sec:emma_policy_guided_optimization}.

\paragraph{Conditioned Observability Encoding.}
\textsc{PolicyDecide} uses compact target-conditioned observability features rather than raw point clouds as scoring input.
For each vehicle-target pair, the feature captures visibility, occlusion, distance, and bearing information derived from the point cloud and LiDAR geometry.
These features describe how well each vehicle can observe the current operator target state from the viewpoint of its LiDAR sensor.

For each vehicle–target pair, we construct a 7-dimensional feature vector $\mathbf{f}=[v,o,p,g,r,\cos\phi,\sin\phi]$ which encodes the target’s observability and relative geometry from the corresponding vehicle’s LiDAR viewpoint. The first four components are ray-based observability ratios, while the remaining three components describe the target distance and bearing in the vehicle-local coordinate system.
Specifically, a LiDAR point cloud can be viewed as a set of returned laser rays, where each point records the depth at which a ray first hits a surface. To reason about target visibility, we use range-view projection to associate each returned point with its corresponding LiDAR ray and observed depth. Given a 3D target box, we select, among the returned rays in the point cloud, those whose directions intersect the box and evaluate only these rays for target observability. For each selected ray, we compare its observed depth with the depth interval where the ray intersects the target box.
If the observed depth falls inside the target interval, the ray is counted as a hit ray.
If the observed depth is smaller than the near side of the target interval,the observed return occurs before the ray reaches the target and is counted as foreground-occluded.
If the observed depth is larger than the far side of the target interval, the ray passes through the target region and is counted as passthrough.
We additionally compute the ratio of rays attributed to ground returns or box-boundary artifacts, allowing the scoring model to distinguish these effects from foreground occlusion.
Let $\mathcal{R}$ be the considered set of rays.
For each ray $i\in\mathcal{R}$, let $z_i$ be the observed depth and $[z_i^{\mathrm{near}}, z_i^{\mathrm{far}}]$ be the target depth interval along this ray.
The four ray-based ratios are computed as normalized counts, where $\mathcal{R}_{\mathrm{gb}}$ denotes rays attributed to ground or boundary artifacts:
\[
\begin{aligned}
v &= \frac{1}{|\mathcal{R}|}\sum_{i\in\mathcal{R}}\mathbb{I}(z_i^{\mathrm{near}}\le z_i \le z_i^{\mathrm{far}}),
& o &= \frac{1}{|\mathcal{R}|}\sum_{i\in\mathcal{R}}\mathbb{I}(z_i < z_i^{\mathrm{near}}),\\
p &= \frac{1}{|\mathcal{R}|}\sum_{i\in\mathcal{R}}\mathbb{I}(z_i > z_i^{\mathrm{far}}),
& g &= \frac{|\mathcal{R}_{\mathrm{gb}}|}{|\mathcal{R}|}.
\end{aligned}
\]
Distance and bearing complement the ray-based observability features. Distance reflects detection difficulty and the potential performance gain from joint perception, while bearing describes the relative viewing direction between the vehicle and the target. In occlusion scenarios, helper vehicles viewing the target from different directions may observe targets that are not visible to the ego vehicle.

For retrieval execution, the current query state may contain multiple target boxes, so each vehicle obtains one observability feature vector for each target. For tracking execution, the current target box is predicted from the historical trajectory state and motion model; EMMA then computes the same 7-dimensional observability feature vector for each candidate vehicle.

\paragraph{Temporal State Feature Encoding.}
\textsc{PolicyTemporal} uses a compact feature vector to summarize how well the current detections match the query targets. For retrieval execution, this temporal state is represented by a detection-to-query matching feature. Given the current detection boxes $\widehat{\mathbf{B}}t$ and the predicate targets $\mathbf{B}^{q}$, EMMA computes their matching distances and counts the query targets matched under the strict and near matching thresholds:
\[
\mathbf{z}_t =
[
d_{\min},
d_{\mathrm{2nd}},
d_{\mathrm{mean}},
d_{\max},
d_{\mathrm{gap}},
n_{\mathrm{strict}},
n_{\mathrm{near}},
r_{\mathrm{strict}},
r_{\mathrm{near}},
N_q
].
\]
The distance terms summarize how closely the current detections match the query targets. The strict and near matching counts and ratios indicate how close the current frame is to satisfying the query, while $N_q$ records the number of query targets.

For tracking execution, the temporal state is represented by a trajectory-conditioned feature vector for skip-length selection:
\[
\mathbf{g}_t =
[
v_1,v_2,v_3,v_4,
a_1,a_2,
s_{\mathrm{vis}},
o_{\mathrm{vis}},
r_{\mathrm{back}}
].
\]
Here, $v_1$ to $v_4$ encode the recent speed history, $a_1$ and $a_2$ encode the recent acceleration history, $s_{\mathrm{vis}}$ and $o_{\mathrm{vis}}$ describe the visibility score and occlusion ratio of the predicted target box, and $r_{\mathrm{back}}$ denotes the recent backjump rate. Stable and predictable motion supports longer skips, while high visibility and low occlusion support reliable target association at the landing frame. In contrast, a high recent backjump rate favors shorter skips.

\section{Policy Training Details}
\label{app:policy_training_details}

\subsection{Oracle References}

We first define operator-specific oracle references for supervision.
For retrieval execution, EMMA defines a target set of frames for which at least one candidate perception configuration, either independent or joint, produces a strict hit.
We define this target set as the oracle strict-hit frame set:
\[
\mathcal{G}_q
=
\left\{
t \in \mathcal{T}
\;\middle|\;
\exists c \in \mathcal{C},
\ \Phi(q, \mathbf{B}_t^{c}) = 1
\right\},
\]
where $\mathcal{C}$ denotes the set of candidate independent and joint perception configurations, $\mathbf{B}_t^{c}$ denotes the detection result produced by candidate $c$ at frame $t$, and $\Phi(\cdot)$ is the predicate matching function.
A strict hit requires every query target to be matched to a distinct detected box within a 3D center-distance threshold of $0.5$ m.

For tracking execution, EMMA defines the oracle reference as the longest
target trajectory that can be continuously recovered from the query's
initial frame by exhaustively evaluating candidate perception actions:
\[
\tau^{\text{oracle}}
=
\left\{
(t,\mathbf{b}^{\text{oracle}}_t)
\mid
t\in\mathcal{T}^{\text{oracle}}
\right\},
\]
where $\mathbf{b}^{\text{oracle}}_t$ denotes the recovered box of the
query target at timestamp $t$, and
\[
\mathcal{T}^{\text{oracle}}
=
\left\{
t_q,t_q+1,\ldots,t_{\mathrm{end}}^{\text{oracle}}
\right\}
\]
denotes the longest consecutive timestamp range starting from the query's initial frame $t_q$, over which the target can be continuously recovered and associated with the trajectory by exhaustively evaluating candidate perception actions.

\subsection{Data Labeling}
All learning-based policies are trained with offline oracle labels.
We first process each training scene with all single-vehicle and collaborative actions, and then match the resulting detections with the ground-truth reference to construct policy-specific supervision.

\paragraph{Retrieval \textsc{PolicyDecide}.}
For retrieval execution, \textsc{PolicyDecide} constructs training frames with respect to the oracle reference set $\mathcal{G}_q$.
Each frame in $\mathcal{G}_q$ satisfies the query predicate and can be detected by at least one candidate perception source.
For each reference frame $g\in\mathcal{G}_q$, we take frames within a bounded temporal neighborhood $\mathcal{N}_{\Delta}(g)$ as candidate training frames, so that the policy observes states near targeted scenes.
We also sample frames that are temporally distant from target scenes as negative examples, maintaining an approximate positive-to-negative ratio of $4{:}1$.

For each candidate scene, let $r(a)$ denote the reward assigned to action $a$.
We partition the candidate frame set into three cases.
(1) \texttt{ind\_hit}: the independent action already achieves an all-strict hit.
We set $r(a_{\mathrm{ind}})=1$ and $r(a_{\mathrm{joint},h})=0$ for all helpers $h$.
(2) \texttt{joint\_needed}: the \texttt{Ind} action misses, but at least one \texttt{Joint} action achieves an all-strict hit.
Successful helpers are ranked by detection quality and assigned
\[
r(a_{\mathrm{joint},h}) = 1 - 10^{-4}\cdot \mathrm{rank}(h),
\]
where $\mathrm{rank}(h)=0$ denotes the best helper; all missed actions receive reward $0$.
(3) \texttt{all\_miss}: all actions miss.
We set $r(a_{\mathrm{ind}})=1$ and $r(a_{\mathrm{joint},h})=0$ for all helpers $h$, so that the policy selects the cheapest action when collaboration provides no benefit.
The actions are ranked according to these rewards and used to construct the training loss.

\paragraph{Retrieval \textsc{PolicyTemporal}.}
For retrieval \textsc{PolicyTemporal}, we compute an oracle reward for every temporal action at each temporal decision state.
For a coarse action $\mathrm{coarse}_k$, we check how many targeted scenes would be skipped among the next $1,\ldots,k-1$ offsets after the current scene.
Let $s_u\in\{0,1\}$ indicate whether scene $u$ is a targeted scene, and let $M_k=\{o\mid 1\le o\le k-1,\ s_{t+o}=1\}$ denote the skipped targeted scenes.
The coarse reward is
\[
r(\mathrm{coarse}_k)
=
\alpha(k-1)
-
\beta \cdot \mathbf{1}[|M_k|>0]
-
\gamma |M_k|,
\]
where $\alpha=0.1$, $\beta=2.0$, and $\gamma=1.0$ by default.
This reward encourages skipping more frames while penalizing skipped targeted scenes.

For a fine action $\mathrm{fine}_r$, the policy searches within a local temporal window of radius $r$ around the current frame $t$.
Using the same targeted-scene indicator $s_u$, we define the left and right targeted sets as
\[
\begin{aligned}
L_r(t)&=\{u\mid u\in[t-r,t),\ s_u=1\},\\
P_r(t)&=\{u\mid u\in(t,t+r],\ s_u=1\}.
\end{aligned}
\]
The fine reward encourages actions that cover targeted scenes in the local temporal window while keeping the window narrow:
\[
r_t(\mathrm{fine}_r)
=
a_L\mathbf{1}[|L_r|>0]
+b_L|L_r|
+a_R\mathbf{1}[|P_r|>0]
+b_R|P_r|
-\lambda r.
\]
We balance the training set so that each temporal action appears as the best action with approximately equal frequency.

\paragraph{Tracking \textsc{PolicyDecide}.}
For tracking execution, \textsc{PolicyDecide} aims to select a perception action that keeps the target continuously and accurately tracked.
For each training target, we generate a teacher trajectory by enumerating all \texttt{Ind} and \texttt{Joint} actions and checking whether the detected object box still matches the ground-truth target.
During trajectory generation, each action receives a teacher reward.
Let $\mathrm{success}(a)$ indicate whether action $a$ keeps the target tracked, and let $d_3(a)$ denote the 3D center distance between the predicted box and the ground truth:
\[
r(a)
=
\mathbf{1}[\mathrm{success}(a)]
+
\mathrm{bonus}(d_3(a))
-
\mathrm{joint\_cost}\cdot \mathbf{1}[\mathrm{joint}(a)].
\]
The bonus term is an accuracy bonus based on $d_3(a)$, rewarding successful actions with more accurate matched detections.
The action with the highest reward is used as the teacher action to update the trajectory and proceed to the next frame.
For final training, each action is converted into a positive or negative label:
\[
y_a=
\mathbf{1}[
\mathrm{matched\_current}(a)
\land
d_3(a)\le 1.5
].
\]

\paragraph{Tracking \textsc{PolicyTemporal}.}
For tracking \textsc{PolicyTemporal}, we label temporal actions by counterfactual skip replay.
A skip action $k$ means that the tracker is propagated by the Kalman filter for $k$ intermediate scenes and then performs one spatial perception update at the landing frame.
For each temporal decision state, we enumerate all candidate skip actions.
After KF-only propagation, the trained \textsc{PolicyDecide} is applied at the landing frame to select a spatial perception action.
The skip is labeled safe if the landing update keeps the tracker matched to the current target and the selected detection is within $1.0\,\mathrm{m}$ 3D center distance to the ground-truth target box:
\[
\mathrm{safe}_t(k)
=
\mathbf{1}\left[
\mathrm{matched\_current}(t_{\mathrm{land}})
\land
d_3(b^{\mathrm{det}}_{t_{\mathrm{land}}}, b^{\mathrm{gt}}_{t_{\mathrm{land}}})
\le 1.0
\right].
\]
Here, $t_{\mathrm{land}}$ is the landing frame after skip action $k$, $b^{\mathrm{det}}_{t_{\mathrm{land}}}$ denotes the selected detection box at the landing frame, and $b^{\mathrm{gt}}_{t_{\mathrm{land}}}$ denotes the ground-truth target box.
This produces a binary safe-label vector over candidate skip actions, and the largest action marked as safe is used as the oracle class label.

\subsection{Training Process}
EMMA trains the policies separately for retrieval and tracking, because the dependency between spatial and temporal decisions differs across the two operators.

For retrieval, EMMA trains \textsc{PolicyTemporal} before \textsc{PolicyDecide}.
The temporal oracle can be derived directly from the oracle reference set $\mathcal{G}_q$, since whether nearby frames should be opened depends on whether they can produce strict-hit results.
We therefore first compute reward vectors for temporal actions and train the temporal action-value model.
We then train \textsc{PolicyDecide} on retrieval decision states by enumerating independent and collaborative actions and assigning oracle labels according to strict-hit quality and execution cost.

For tracking, EMMA trains \textsc{PolicyDecide} before \textsc{PolicyTemporal}.
Unlike retrieval, tracking is closed-loop: a temporal skip changes the tracker state and affects future spatial decisions.
EMMA therefore first trains \textsc{PolicyDecide} from teacher rollouts without skipping any scene, so that the policy learns which spatial perception action preserves target matching at each tracking state.
After freezing the trained \textsc{PolicyDecide}, EMMA runs closed-loop rollouts to derive safe-skip labels, which are then used to train \textsc{PolicyTemporal}.

\subsection{Loss Function Design}
Retrieval \textsc{PolicyTemporal} is trained as a regression model, while tracking \textsc{PolicyTemporal} and both \textsc{PolicyDecide} models are trained with classification or ranking objectives.

\paragraph{Retrieval \textsc{PolicyDecide}.}
Let $s_a$ be the model score for action $a$, and let $a^*$ be the oracle best action.
The training objective is
\[
\mathcal{L}
=
\alpha_{CE}\mathcal{L}_{CE}
+
\alpha_{rank}\mathcal{L}_{rank}
+
\mathcal{L}_{gate}
+
\mathcal{L}_{multi\text{-}pos}.
\]
The cross-entropy term encourages the oracle best action to receive the highest score:
\[
\mathcal{L}_{CE}
=
-\log
\frac{\exp(s_{a^*})}{\sum_a \exp(s_a)}.
\]
The ranking term separates the best action from lower-reward actions by margin $m$:
\[
\mathcal{L}_{rank}
=
\sum_{a:r(a)<r(a^*)}
\max(0,\ m-(s_{a^*}-s_a)).
\]
The gate loss learns when to switch between independent and collaborative actions based on oracle rewards.
If $\max_h r(a_{\mathrm{joint},h})-r(a_{\mathrm{ind}})\ge 0.5$, the collaborative score is encouraged to be higher than the independent score; otherwise, the independent score is encouraged to be higher.
Let $P$ and $N$ denote the positive and negative helper sets.
The multi-positive helper loss allows multiple helpers to be positive and encourages positive helpers to score above negative helpers:
\[
\mathcal{L}_{multi\text{-}pos}
=
\frac{1}{|P||N|}
\sum_{p\in P,n\in N}
\max(0,\ m-(s_p-s_n)).
\]

\paragraph{Retrieval \textsc{PolicyTemporal}.}
Retrieval \textsc{PolicyTemporal} is an MLP action-value scorer trained to regress the oracle reward vector.
For sample $i$ and temporal action $a$, let $r_{i,a}$ be the oracle coarse/fine reward, $\hat{r}_{i,a}$ be the predicted reward, and $w_{i,a}$ be the training weight.
The loss is
\[
\mathcal{L}_{temporal}
=
\frac{1}{N}
\sum_i\sum_a
w_{i,a}
(\hat{r}_{i,a}-r_{i,a})^2.
\]
At inference time, EMMA chooses the temporal action with the largest predicted reward.

\paragraph{Tracking \textsc{PolicyDecide}.}
Tracking \textsc{PolicyDecide} is a role-aware action scorer trained with layout dropout.
The training loss is
\[
\mathcal{L}
=
1.0\mathcal{L}_{hitBCE}
+
0.3\mathcal{L}_{setRank}
+
0.5\mathcal{L}_{hardNeg}.
\]
Here, $\mathcal{L}_{hitBCE}$ trains action-level hit prediction using positive and negative labels, $\mathcal{L}_{setRank}$ requires at least one positive action to score higher than negative actions, and $\mathcal{L}_{hardNeg}$ suppresses the highest-scoring hard negative actions.

\paragraph{Tracking \textsc{PolicyTemporal}.}
Tracking \textsc{PolicyTemporal} is an MLP classifier over skip actions.
It is trained with cross-entropy loss:
\[
\mathcal{L}_{temp}
=
-\log
\frac{\exp(\mathrm{logit}_{y_i})}
{\sum_j \exp(\mathrm{logit}_j)}.
\]
At inference time, EMMA chooses the skip action with the largest logit.

\subsection{Tracking Policy Training Curves}
\label{app:tracking_training_curves}
Fig.~\ref{fig:tracking_training_curves} reports the training curves of the two tracking policies.
\begin{figure}[ht]
    \centering
    \begin{subfigure}[t]{0.48\linewidth}
        \centering
        \includegraphics[width=\linewidth]{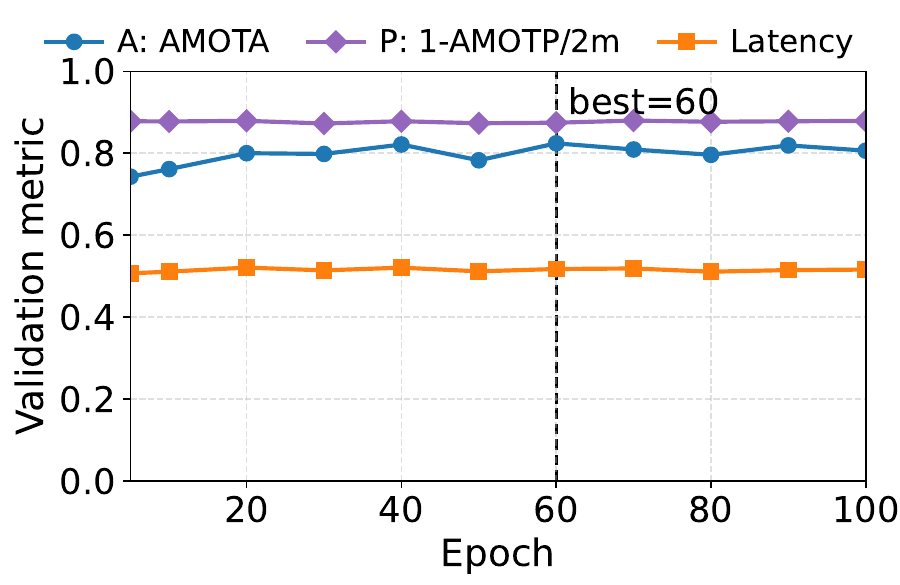}
        \caption{PolicyDecide}
        \label{fig:trackingdecide}
    \end{subfigure}
    \hfill
    \begin{subfigure}[t]{0.48\linewidth}
        \centering
        \includegraphics[width=\linewidth]{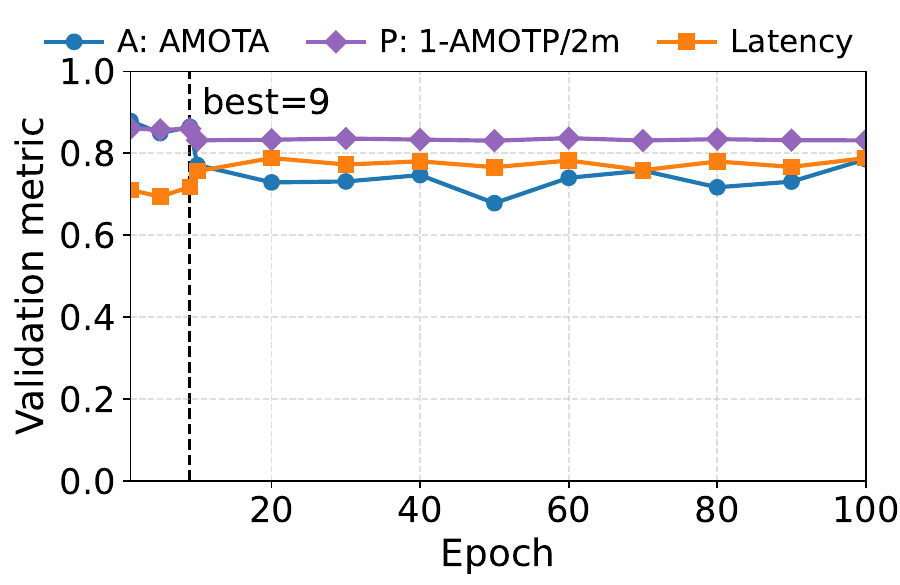}
        \caption{PolicyTemporal}
        \label{fig:trackingtemporal}
    \end{subfigure}
    \vspace*{-2mm}
    \caption{Tracking policy training curves.}
    \label{fig:tracking_training_curves}
\end{figure}

\end{document}